\documentclass[twocolumn]{aastex62}
\usepackage{natbib}
\usepackage{xcolor}
\usepackage[caption=false]{subfig}
\hypersetup{colorlinks,,linkcolor={red},citecolor={blue},urlcolor={blue}, 
 pdflang={English}}
\shorttitle{Radio \& optical study of 3C 454.3}
\shortauthors{Sarkar et al.}

\begin{document}

\title{Long term variability and correlation study of the blazar 3C 454.3 in radio, NIR and optical wavebands}

\correspondingauthor{A. Sarkar}
\email{sarkadipta@gmail.com, arkadipta.sarkar@tifr.res.in}

\author{A. Sarkar}
\affiliation{Department of High Energy Physics, Tata Institute of Fundamental Research, Mumbai, 400005, India}

\author{V. R. Chitnis}
\affiliation{Department of High Energy Physics, Tata Institute of Fundamental Research, Mumbai, 400005, India}

\author{A. C. Gupta}
\affiliation{Aryabhatta Research Institute of Observational Sciences (ARIES), Manora Peak, Nainital 263002, India}

\author{H. Gaur}
\affiliation{Aryabhatta Research Institute of Observational Sciences (ARIES), Manora Peak, Nainital 263002, India}

\author{S. R. Patel}
\affiliation{Department of High Energy Physics, Tata Institute of Fundamental Research, Mumbai, 400005, India}
\affiliation{Department of Physics, University of Mumbai, Santacruz (E), Mumbai-400098, India}

\author{P.  J. Wiita}
\affiliation{Department of Physics, The College of New Jersey, PO Box 7718, Ewing, NJ 08628-0718, USA}

\author{A.E. Volvach}
\affiliation{Radio Astronomy Laboratory of Crimean Astrophysical Observatory, Katsively, RT-22 Crimea}
\affiliation{Astro Space Center, Lebedev Physical Institute, Russian Academy of Sciences, Profsoyuznaya ul. 84/32, Moscow, 117997 Russia}

\author{M. Tornikoski}
\affiliation{Aalto University Mets\"ahovi Radio Observatory, Mets\"ahovintie 114, 02540 Kylm\"al\"a, Finland}

\author{W. Chamani}
\affiliation{Aalto University Mets\"ahovi Radio Observatory, Mets\"ahovintie 114, 02540 Kylm\"al\"a, Finland}
\affiliation{Aalto University Department of Electronics and Nanoengineering, P.O. Box 15500, FI-00076 Aalto, Finland}

\author{S. Enestam}
\affiliation{Aalto University Mets\"ahovi Radio Observatory, Mets\"ahovintie 114, 02540 Kylm\"al\"a, Finland}
\affiliation{Aalto University Department of Electronics and Nanoengineering, P.O. Box 15500, FI-00076 Aalto, Finland}

\author{A. L\"ahteenm\"aki}
\affiliation{Aalto University Mets\"ahovi Radio Observatory, Mets\"ahovintie 114, 02540 Kylm\"al\"a, Finland}
\affiliation{Aalto University Department of Electronics and Nanoengineering, P.O. Box 15500, FI-00076 Aalto, Finland}

\author{J. Tammi}
\affiliation{Aalto University Mets\"ahovi Radio Observatory, Mets\"ahovintie 114, 02540 Kylm\"al\"a, Finland}

\author{R.J.C Vera}
\affiliation{Aalto University Mets\"ahovi Radio Observatory, Mets\"ahovintie 114, 02540 Kylm\"al\"a, Finland}
\affiliation{Aalto University Department of Electronics and Nanoengineering, P.O. Box 15500, FI-00076 Aalto, Finland}

\author{L.N. Volvach}
\affiliation{Radio Astronomy Laboratory of Crimean Astrophysical Observatory, Katsively, RT-22 Crimea}
\affiliation{Astro Space Center, Lebedev Physical Institute, Russian Academy of Sciences, Profsoyuznaya ul. 84/32, Moscow, 117997 Russia}

\begin{abstract}
We performed a long-term optical (B, V, R bands), infra-red (J and K bands) and radio band (15, 22, 37 GHz band) study on the flat spectrum radio quasar, 3C 454.3, using the data collected over a period of more than 8 years (MJD 54500--57500). The temporal variability, spectral properties and inter-waveband correlations were studied by dividing the available data into smaller segments with more regular sampling. This helped us constrain the size and the relative locations of the emission regions for different wavebands. Spectral analysis of the source revealed the interplay between the accretion disk and jet emission. The source predominantly showed a redder-when-brighter trend, though we observed a bluer-when-brighter trend at high flux levels which could be signatures of particle acceleration and radiative cooling. Significant correlations with near-zero lag were seen between various optical/infra-red bands, indicating that these emission regions are co-spatial. Correlations with a time lag of about 10--100 days are seen between optical/infra-red and radio bands indicating these emissions arise from different regions. We also observe the DCF peak lag change from year to year. We try to explain these differences using a curved jet model where the different emission regions have different viewing angles resulting in a frequency dependent Doppler factor. This variable Doppler factor model explains the variability timescales and the variation in DCF peak lag between the radio and optical emissions in different segments. Lags of 6-180 days are seen between emissions in various radio bands, indicating a core-shift effect.
\end{abstract}

\keywords{galaxies: active --- galaxies :jet --- methods: observational --- quasars: individual (3C 454.3) --- techniques: photometric}

\section{Introduction}
\label{sec:1}

Blazars are a subclass of Active Galactic Nuclei (AGN) whose relativistic jets point approximately towards our line of sight \citep{UrryPadovani1995}. They are further divided into two sub-classes: Flat Spectrum Radio Quasars (FSRQ) and BL Lacertae like objects (BL Lacs). FSRQs show strong emission lines but BL Lacs do not show any significant absorption or emission lines. 3C 454.3 is an FSRQ at a redshift of $z_{rs}=0.859$ and is a highly variable source. 

3C 454.3 has very peculiar multi-wavelength variability properties and because of that, it has been subjected to several simultaneous multi-waveband observational  campaigns. In the spring of 2005, an exceptional outburst was detected in this blazar in all wavebands from mm to X-rays \citep[see][]{2006A&A...456..911G, 2006A&A...445L...1F, 2006A&A...449L..21P}. As a follow up, several multi-wavelength campaigns were carried out \citep[e.g.,][]{2006A&A...453..817V, 2007A&A...464L...5V, 2007A&A...473..819R, 2008A&A...491..755R, 2008A&A...485L..17R}.

After the launch of the \textit{Fermi} satellite, it was recognized that 3C 454.3 is one of the brightest $\gamma-$ray emitting blazars \citep{2010ApJ...721.1383A}. This blazar is listed as 1FGL J2253.9$+$1608 in the First \textit{Fermi} Large Area Telescope (LAT) AGN catalog \citep{2010ApJ...722..520A}. During multi-waveband observations of this blazar carried out during  2008 August to December, \citet{Bonningetal2009} found excellent correlations between near-infrared (NIR), optical, UV and $\gamma-$ray fluxes, with a time lag of less than one day. However, the X-ray flux was almost non-variable and not correlated with either the higher or lower frequency measurements. \citet{Vercelloneetal2009} noticed correlated optical and high energy $\gamma-$rays measurements by \textit{AGILE} in 2007 November observations. However, the X-ray observations from \textit{INTEGRAL} and \textit{Swift}-XRT were not correlated with optical/$\gamma-$ ray flux emissions. In a more complete \textit{AGILE} led multi-waveband monitoring of 3C 454.3 during 2008 May to 2009 January, \citet{2010ApJ...712..405V} found nearly simultaneous flux peaks across all bands during the strong flares, with the $\gamma-$optical correlation having a time lag of less than a day. Strong correlations between $\gamma-$ray and optical light curves (LCs) were found by \citet{2012AJ....143...23G} during a 2009 November -- December flare, though in this case, the optical emission led the $\gamma-$rays by 4.5$\pm$1.0 days. The X-ray LC was essentially constant and hence showed no correlation with the other bands. Similar strong correlations were found between NIR-optical and $\gamma-$rays by \citet{2017MNRAS.464.2046K}, with optical-NIR emission leading \textit{Fermi}-LAT $\gamma-$rays emission by $\sim$ 3 days in observations taken from 2014 October 19 to 2014 December 23. \citet{2017MNRAS.464.2046K} also noticed that optical-NIR and $\gamma-$ray emissions were well correlated without any delay in three different time slots. A very peculiar behavior is seen in the blazar flare observed during 2009 December 3 --12 in which $\gamma-$ray, X-ray, optical, and NIR fluxes peaked nearly at the same time with optical polarization showing dramatic changes during the flare whereas cm-band radio data showed no correlation with variations at higher frequencies. However, there was a strong anticorrelation between optical flux and degree of polarization along with large, rapid swings in polarization angle of $\sim$ 170$^\circ$ \citep{2017MNRAS.472..788G}.

The very peculiar nature of 3C 454.3 motivated us to perform a detailed multi-wavelength study of the source for an extended period of observation. Main goal of our study is to shed light upon the emission mechanism and the origin of variability. In the present work, we focused on simultaneous multi-wavelength long-term low energy observations of this FSRQ taken during 2008 February to 2016 April. We searched them for variability and spectral properties, and determined inter-waveband cross correlations.

Being a very well studied source, our data partially overlap with those of a few studies conducted in the past. For the observational period of 15 April 2009 to 1 August 2011, a detailed multi-wavelength study showed three prominent $\gamma-$ray outbursts: 2009 Autumn, 2010 Spring, and 2010 Autumn as seen by \citet{Jorstadetal2013}. They explained the multi-waveband behavior using a system of standing conical shocks along with magnetic reconnection events in the millimeter waveband core of the jet. Our R, J, K and 37 GHz band data overlaps with this study from April 2009 to August 2011 though our motivation is very different. For the observation period, June 2007 to January 2010, \citet{Kutkinetal2014} analyzed multifrequency radio band data to study the core shift effect in 3C 454.3. Our 22 and 37 GHz band data are common with them for the period of February 2008 to December 2009. \citet{Kutkinetal2014} have analyzed two strong radio flares during this period and studied multi-frequency radio band cross correlation. We however, have studied cross-correlation only for the first flare and our results differ from those obtained by \citet{Kutkinetal2014}. \citet{2016MNRAS.456..171R} analyzed multifrequency $\gamma-$ray, optical and radio band data of 15 blazars including 3C 454.3. They studied cross correlation between R vs 37 GHz radio data for entire data stretch. During this period (2012.5 to 2015), the source underwent a significant increase in flux density (from 4 Jy tp 20 Jy) in the 37 GHz band, and there were substantial optical flaring activities (see Fig.\ 1). However, the increase in mm band output was very slow and its substructure is not very well-defined, compared to the other (even stronger) flare in 2010-2011. A part of our study (including R and 37 GHz band data) is similar to \citet{2016MNRAS.456..171R}; however, we subdivide the lightcurve to improve the sampling and observe the temporal evolution of the inter-waveband correlations. In the same sense, the present study differs from \citet{2018MNRAS.480.5517L} where they analyzed the correlation between 15 GHz and optical bands during July 2009 to November 2017 and concluded that the optical emissions led the radio emissions by $403\pm 6$ days.

In the present work, the observatories and the data acquisition methodology are discussed briefly in section \ref{sec:2}. The long-term multi-waveband light curves for the
optical/IR and radio bands are presented in section \ref{sec:6}, followed by the variability study in section \ref{sec:7} and the study of spectral variations in \ref{sec:8}.
Section \ref{sec:9} analyzes the correlations between different wave-bands. This is followed by a discussion section \ref{sec:10} where we try to estimate the emission region sizes
for different wavebands and model the emission using a frequency dependant Doppler factor model. Our conclusions follow in section \ref{sec:11}.

\begin{figure*}[!]
\centering
\includegraphics[width=0.95\linewidth]{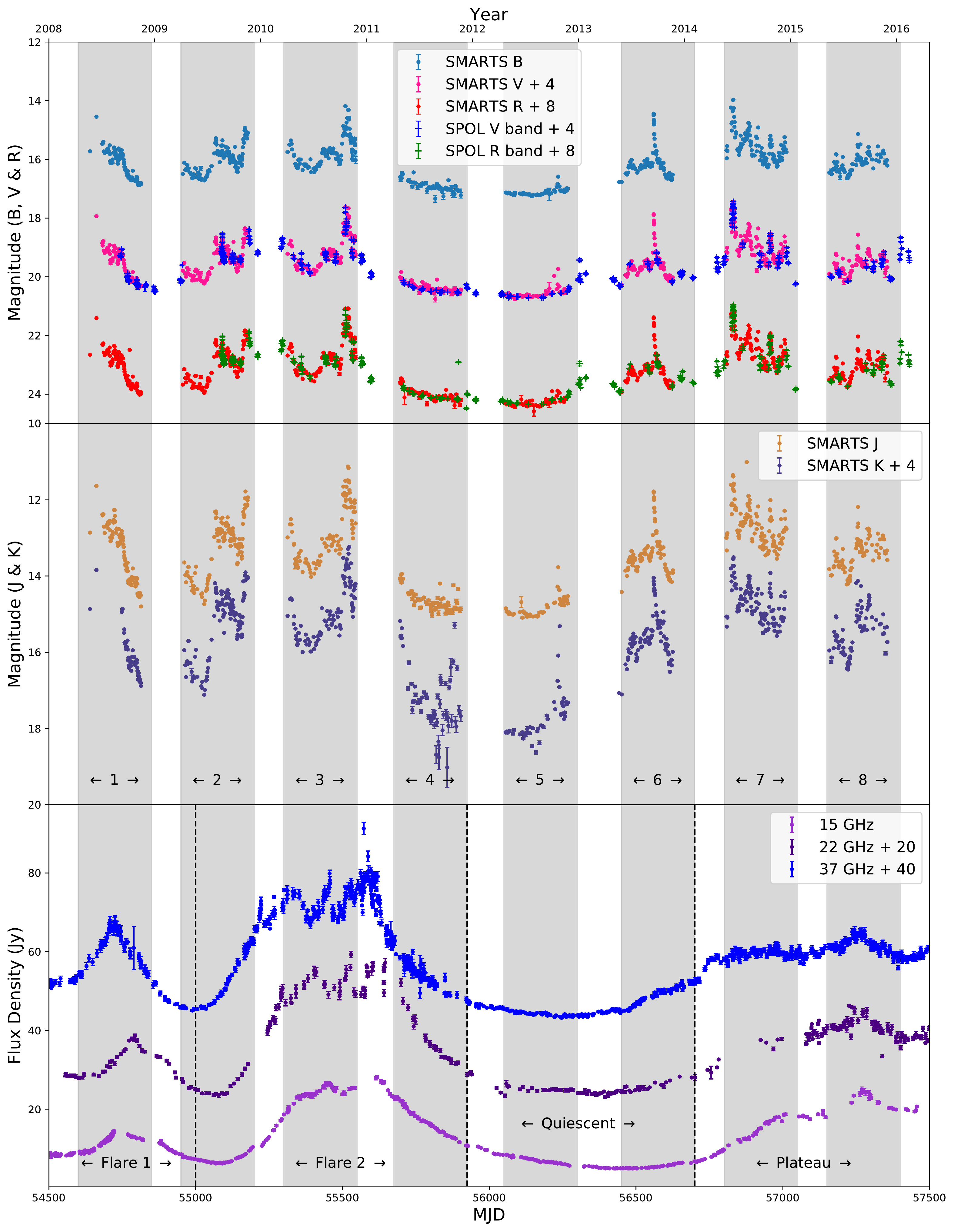}
\caption{\label{f1}
Multiwaveband light curves for 3C 454.3: \textbf{(top)} optical bands; \textbf{(middle)} IR bands; \textbf{(bottom)} radio bands. Curves in different bands are shifted as indicated in the legends to improve visibility and the eight individual segments are shaded in gray. In the last panel, the dotted lines separate the radio light curves into four segments where further analyses were performed.}
\end{figure*}

\section{Observations}
\label{sec:2}
The data we consider span a period of more than eight years from February 2, 2008, through April 22, 2016, in optical, IR and radio bands. Here we briefly discuss the data taken from public archives and the new observations and their analysis carried out by us.

\subsection{Optical and Infra-red data}
\label{sec:3}

The optical and near infra-red (NIR) data for 3C 454.3 are taken from the public archive of the Small and Moderate Aperture Research Telescope System (SMARTS) and Steward Observatory telescopes. SMARTS is a part of the Cerro Tololo Inter-American Observatory (CTIO) and has been observing all \textit{Fermi}-Large Area Telescope (LAT) monitored blazars in optical B, V, R bands and NIR J and K bands that are accessible from Chile. Details about the SMARTS telescopes, detectors, observations, and data analysis are given in \citet{Bonningetal2012, buxton12_optic_near_infrar_monit_black}. Steward Observatory of the University of Arizona uses the 2.3m Bok and 1.54m Kuiper telescopes to carry out optical photometric and polarimetric observations of a large number of blazars using the spectropolarimeter (SPOL). We have taken the optical photometric data in V and R bands of the blazar 3C 454.3 from their archive. Details about these telescopes, instruments, observations and data analysis methods are given in \citet{Smithetal2009}.

\subsection{Radio Data}
\label{sec:4}

Observations at 37 GHz were made with the 14-m radio telescope of Aalto University Mets{\"a}hovi Radio Observatory. The detection limit of the telescope at 37 GHz under optimal conditions is of the order of 0.2 Jy. Data points with a signal-to-noise ratio $<$ 4 are handled as non-detections. The error estimate in the flux density includes the contribution from the measurement rms and the uncertainty of the absolute calibration. A detailed description of the data reduction and analysis of Mets{\"a}hovi data is given in \citet{Teraesrantaetal1998}. From 2008 to mid-2016 933 data points were obtained at 37 GHz for 3C 454.3, with data taken at mean intervals of 3.3 days, median intervals of 1.9 days, and with a maximum gap between two data points being 32 days.

Observations at 22 and 37 GHz were performed using the 22-m radio telescope (RT-22) of the Crimean Astrophysical Observatory (CrAO).  The source signal was determined as the difference between the radiometer responses at the two antenna positions (source and the sky), averaged over 30 sec. Depending on the source flux density, 5--20 measurements were performed and the corresponding mean values and rms errors were determined. The effective area of the 22-m telescope was determined from observations of calibrator sources. The uncertainties due to equipment noise, telescope pointing errors, atmospheric absorption errors, and instability of the radiometer gain were taken into account while calculating the rms error in flux density. Details about the CrAO telescope, observation techniques and data analysis are discussed in \citet{Nesterovetal2000, Volvach2006}. In our analyses, the Mets{\"a}hovi and CrAO data at 37 GHz are combined. The mean cadence for the combined 37 GHz data is 1.3 days.

Observations at 15 GHz were taken by the 40 m Owens Valley Radio Telescope (OVRO). The telescope observes over 1500 northern ($>20^0$) sources from the Candidate Gamma-ray Blazar Survey (CGRaBS).  Each source is observed twice a week with a $\sim 4$ mJy (minimum) and 3\% (typical) uncertainties. The detailed working of the telescope is presented in \citet{Richardsetal2011}.

\section{Analysis and results}
\label{sec:5}

We use the optical, IR and radio data to plot the long-term multi-waveband light curve of 3C 454.3. We also study the variability, spectral properties, and correlations between various wavebands.

\subsection{Light Curves}
\label{sec:6}

The multi-waveband light curves, which span more than eight years, are given in Figure \ref{f1}.  The figure shows data of B, V, R, J, K bands in the optical/IR and 15 GHz, 22 GHz and 37 GHz in the radio from MJD 54500 to MJD 57500 (Feb 2008 to Apr 2016). We have examined annual changes in variability, spectral properties and the correlations between various wavebands. For this purpose, we subdivided the available light curves into eight segments of 250 days each (Segments 1--8, in Figure \ref{f1}) based on the availability of optical/IR data. We also subdivided the radio light curves into four different segments based on the activity of the source. The first segment includes the October 2008 flare, the second segment includes the complex long-term flare from June 2009 to December 2011. The third segment is when the source was in a quiescent state from January 2012 to February 2014. The final segment consists of the rest of the light curve where the flux density is higher than in the quiescent state but lower than in the flaring state. The start and stop dates of the individual segments are given in Table \ref{t0}.

The series of analyses that are performed in subsequent sections are very sensitive to outliers. It is therefore essential to properly identify and remove outlier points in the data. Two methods were used to identify outliers. Firstly, for optical/IR bands where there are measurements carried out on same day in different bands, a scatter plot of fluxes from nearby bands from same observatory was constructed and the best fit line was obtained. Any point that lies beyond $3 \sigma$ from the best fit line is identified as an outlier, where $\sigma$ is the standard deviation about the best-fit line. This ensures similar variations in nearby wavebands, which are expected since the emission mechanism is similar. Next, for radio wavebands, where we do not always have data collected on same day at different frequencies, we take the distribution of $K_i = \frac{|\mathcal{F}_{i+1} - \mathcal{F}_i|}{\Delta_{i+1,i}}$, where $\mathcal{F}_i$ and $\mathcal{F}_{i+1}$ are the flux densities from i-th and (i+1)-th observation and $\Delta_{i+1,i}$ is the time difference between i-th and (i+1)-th observation. Points where $K_i$ and $K_{i-1}$ are more than three standard deviations away from the mean are identified as outliers. Here it  is assumed that a large and sudden change in flux density is unlikely.

\begin{deluxetable}{ccCc}[!h]
\tablecaption{Start and stop dates of different segments.\label{t0}}
\tablecolumns{4}
\tablenum{1}
\tablewidth{0pt}
\tablehead{
\colhead{Seg\tablenotemark{1}} & \colhead{Start Time} & \colhead{Stop Time } & \colhead{Length} \\
\colhead{}    & \colhead{MJD}        & \colhead{MJD}        & \colhead{(days)}
}
\startdata
      1 & 54600 & 54850 & 250 \\
      2 & 54950 & 55200 & 250 \\
      3 & 55300 & 55550 & 250 \\
      4 & 55675 & 55925 & 250 \\
      5 & 56050 & 56300 & 250 \\
      6 & 56450 & 56700 & 250 \\
      7 & 56800 & 57050 & 250 \\
      8 & 57150 & 57400 & 250 \\
      \hline
      Flare 1   & 54500 & 55000 & 500 \\
      Flare 2   & 55000 & 55925 & 925 \\
      Quiescent & 55925 & 56700 & 775 \\
      Plateau   & 56700 & 57500 & 800 \\ 
      \enddata
\tablenotetext{1}{Segments 1--8 are based on the annual availability of the optical data; the others divide the radio light curves based on the source activity.}
\end{deluxetable}

\subsection{Variability Study}
\label{sec:7}
In this section, we quantify the variations of the source flux in different wavebands. Variability can be crudely divided into three types depending on their timescales: intra-day (IDV), where the variations occur rapidly within a single day; short-term, where the variations occur on timescales of days or weeks; and finally, long-term variability, where the time scale ranges from months to years (or even decades for some sources). Using the optical/IR and radio data, we quantify the variability of the source in different segments using a $\chi^2$ test and variability amplitude parameter $V_F$. $\chi^2$ is defined as:

\begin{equation}
  \chi^2 = \sum_{i=1}^N\frac{(\mathcal{F}_i-\bar{\mathcal{F}})^2}{\mathcal{E}_i^2},
\end{equation}

where $\mathcal{F}_i$ is the flux density at the $i^{th}$ observation and $\mathcal{E}_i$ is the corresponding error. The resulting $\chi^2$ value is then compared with a critical value which depends on the number of degrees of freedom (i.e., number of observations minus the number of free parameters) and the level of significance required. If the obtained $\chi^2$ value is more than the critical value, the source is considered as variable at the corresponding significance level. For all our tests we used a significance level of 99.9\%. These $\chi^2$ tests revealed the source to be variable in all segments (and in the entirety) and in all wavebands (optical/IR and radio)  at that level. We also estimated the variability amplitude \citep{HeidtWagner1996}. It is defined as:

\begin{equation}
  V_F = (1/ \bar{\mathcal{F}})\sqrt{(\mathcal{F}_{max} - \mathcal{F}_{min})^2-2 \bar{\mathcal{E}}^2},
\end{equation}

where $\mathcal{F}_{min/max}$ corresponds to the minimum/maximum flux density in the light curve, $\bar{\mathcal{F}}$ is the average flux and $\bar{\mathcal{E}}$ is the average uncertainty in the flux measurements. These statistical tools help to quantify the variability of the source in different wavebands. The timescale of variability was estimated using the flux doubling/halving timescales ($t_d$):

\begin{equation}
  \mathcal{F}(t_{i+1}) = \mathcal{F}(t_i) 2^{(t_{(i+1)}-t_i)/t_d},
\end{equation}

where $\mathcal{F}(t_i)$ is the flux density at time $t_i$ and $t_d$ is the characteristic timescale of flux doubling \citep{Foschinietal2011}. The shortest flux doubling timescale is used as an estimate for the variability timescale in our calculations ($t_d\approx\tau_{var}$). We compute $V_F$ and $t_d$ for the optical, IR and radio light curves in all the segments.

In case of optical/IR data, only one data point was available for a particular waveband each day. Combining the two observatories, we had at the most, two data points per day. These low statistics meant that we could not explicitly look for IDVs in optical/IR bands.The source was observed to be variable with 99.9\% confidence in all the observed wavebands in all the segments. The variability amplitudes and timescales in the individual segments are given in Table \ref{t2}. Fast flux variability was observed in Segment 6, with $t_d<1$ day bringing these variability in the IDV regime.

\begin{deluxetable*}{cccc|cccc|cccc|cccc}[!htb]
\tablecaption{\label{t2} Optical/IR variability of 3C 454.3 in different segments}
\tablecolumns{16}
\tablenum{2}
\tablewidth{0pt}
\tablehead{
  \colhead{$\Delta_t$} & \colhead{Band}  & \colhead{$V_F$} & \colhead{$t_d$} & \colhead{$\Delta_t$} & \colhead{Band}  & \colhead{$V_F$} & \colhead{$t_d$} &
  \colhead{$\Delta_t$} & \colhead{Band}  & \colhead{$V_F$} & \colhead{$t_d$} & \colhead{$\Delta_t$} & \colhead{Band}  & \colhead{$V_F$} & \colhead{$t_d$} \\
  \colhead{} & \colhead{}  & \colhead{} & \colhead{(days)} & \colhead{} & \colhead{}  & \colhead{} & \colhead{(days)} &
  \colhead{} & \colhead{}  & \colhead{} & \colhead{(days)} & \colhead{} & \colhead{}  & \colhead{} & \colhead{(days)}
}
\startdata
            & B & 3.58  & 1.83 &       & B & 2.98 & 0.81 &        & B   & 0.74 & 9.53 &       & B  & 3.33 & 1.44 \\
            & V & 3.74  & 1.85 &       & V & 2.90 & 0.82 &        & V   & 1.27 & 5.34 &       & V  & 3.14 & 0.70  \\
      Seg 1 & R & 3.87  & 1.91 & Seg 3 & R & 3.04 & 0.82 &  Seg 5 & R   & 0.67 & 7.92 & Seg 7 & R  & 3.10 & 1.46 \\
            & J & 3.90  & 2.04 &       & J & 3.54 & 1.10 &        & J   & 1.77 & 3.02 &       & J  & 3.76 & 0.61 \\
            & K & 4.11  & 2.10 &       & K & 3.58 & 0.88 &        & K   & 2.11 & 1.61 &       & K  & 2.57 & 1.11 \\
      \hline                                                       
            & B & 2.02  & 1.06 &       & B & 0.85 & 4.32 &        & B   & 3.43 & 0.63 &       & B  & 1.79 & 1.54 \\
            & V & 2.00  & 1.12 &       & V & 0.92 & 4.65 &        & V   & 3.45 & 0.60 &       & V  & 1.88 & 1.46 \\
      Seg 2 & R & 2.18  & 1.75 & Seg 4 & R & 0.94 & 4.81 &  Seg 6 & R   & 3.55 & 0.68 & Seg 8 & R  & 2.09 & 1.27 \\
            & J & 2.87  & 1.96 &       & J & 1.16 & 3.65 &        & J   & 3.26 & 0.61 &       & J  & 2.44 & 1.27 \\
            & K & 2.32  & 1.83 &       & K & 1.43 & 3.69 &        & K   & 2.89 & 0.72 &       & K  & 2.88 & 1.12 \\
      \hline     
\enddata
\end{deluxetable*}

\begin{deluxetable*}{cccc|cccc|cccc|cccc}[!htb]
\tablecaption{\label{t3} Radio variability of 3C 454.3 in different segments}
\tablecolumns{16}
\tablenum{3}
\tablewidth{0pt}
\tablehead{
  \colhead{$\Delta_t$} & \colhead{$\nu$}  & \colhead{$V_F$} & \colhead{$t_d$ } &   \colhead{$\Delta_t$} & \colhead{$\nu$}  & \colhead{$V_F$} & \colhead{$t_d$} & 
  \colhead{$\Delta_t$} & \colhead{$\nu$}  & \colhead{$V_F$} & \colhead{$t_d$ } &   \colhead{$\Delta_t$} & \colhead{$\nu$}  & \colhead{$V_F$} & \colhead{$t_d$} \\
  \colhead{} & \colhead{(GHz)}  & \colhead{} & \colhead{(days)} &   \colhead{} & \colhead{(GHz)}  & \colhead{} & \colhead{(days)} & 
  \colhead{} & \colhead{(GHz)}  & \colhead{} & \colhead{(days)} &   \colhead{} & \colhead{(GHz)}  & \colhead{} & \colhead{(days)}
}
\startdata
            & 15  &  0.58 & 19.42 &         & 15 &  1.28 & 67.04 &           & 15  & 0.92  & 37.20 &           & 15  & 1.15 & 56.56 \\ 
    Seg 1   & 22  &  1.17 & 12.29 & Seg 2   & 22 &  1.51 & 9.31  & Seg 3     & 22  & 1.26  & 7.62  & Seg 4     & 22  & 0.93 & 10.41 \\ 
            & 37  &  1.37 & 7.24  &         & 37 &  1.86 & 3.22  &           & 37  & 1.49  & 4.90  &           & 37  & 0.68 & 7.07  \\
\enddata
\tablecomments{Flux doubling timescale is calculated ignoring intra-day observations.}            
\end{deluxetable*}           

\begin{deluxetable*}{ccccc|ccccc|ccccc}[!htb]
\tablecaption{\label{t8} IDV measurements for 15 GHz band.}
\tablecolumns{12}
\tablenum{4}
\tablewidth{0pt}
\tablehead{
  MJD & \colhead{$\chi_r^2$} & \colhead{$\chi^2_{0.999,r}$ } & \colhead{Var?} & \colhead{$t_d$} &
  MJD & \colhead{$\chi_r^2$} & \colhead{$\chi^2_{0.999,r}$ } & \colhead{Var?} & \colhead{$t_d$} &
  MJD & \colhead{$\chi_r^2$} & \colhead{$\chi^2_{0.999,r}$ } & \colhead{Var?} & \colhead{$t_d$} \\
      & \colhead{} & \colhead{} & \colhead{} & \colhead{(days)} &
      & \colhead{} & \colhead{} & \colhead{} & \colhead{(days)} &
      & \colhead{} & \colhead{} & \colhead{} & \colhead{(days)}
}
\startdata
54695  & 2.07 & 2.45 & No  & -     & 54705  & 0.43 & 2.45 & No  & -     & 54714  & 4.07 & 2.74 & Yes & 0.10 \\
54697  & 1.54 & 2.65 & No  & -     & 54706  & 2.36 & 2.51 & No  & -     & 54715  & 1.25 & 2.58 & No  & -    \\
54698  & 2.23 & 2.45 & No  & -     & 54707  & 4.92 & 2.45 & Yes & 0.04  & 54716  & 2.62 & 2.30 & Yes & 0.04 \\
54699  & 0.95 & 2.26 & No  & -     & 54708  & 3.39 & 2.45 & Yes & 9.78  & 54717  & 3.68 & 2.35 & Yes & 0.10 \\
54700  & 1.23 & 2.45 & No  & -     & 54709  & 1.32 & 2.65 & No  & -     & 54718  & 4.03 & 2.74 & Yes & 1.05 \\
54701  & 4.07 & 2.35 & Yes & 0.29  & 54710  & 0.33 & 2.74 & No  & -     & 54719  & 1.74 & 2.51 & No  & -    \\
54702  & 4.92 & 2.58 & Yes & 0.18  & 54711  & 0.38 & 2.51 & No  & -     & 54722  & 1.59 & 2.58 & No  & -    \\
54703  & 6.73 & 2.35 & Yes & 0.06  & 54712  & 2.61 & 2.51 & Yes & 0.06  & 55315  & 1.96 & 2.13 & No  & -    \\
54704  & 1.15 & 2.45 & No  & -     & 54713  & 2.48 & 2.30 & Yes & 0.01  & 55316  & 1.73 & 2.16 & No  & -    \\
\enddata
\tablecomments{MJD 54708: though $\chi^2$ test gives the variability to be significant, $t_d$ is larger than IDV timescales.}
\end{deluxetable*}

\begin{figure*}[!t]
  \centering
  \includegraphics[width=0.95\linewidth]{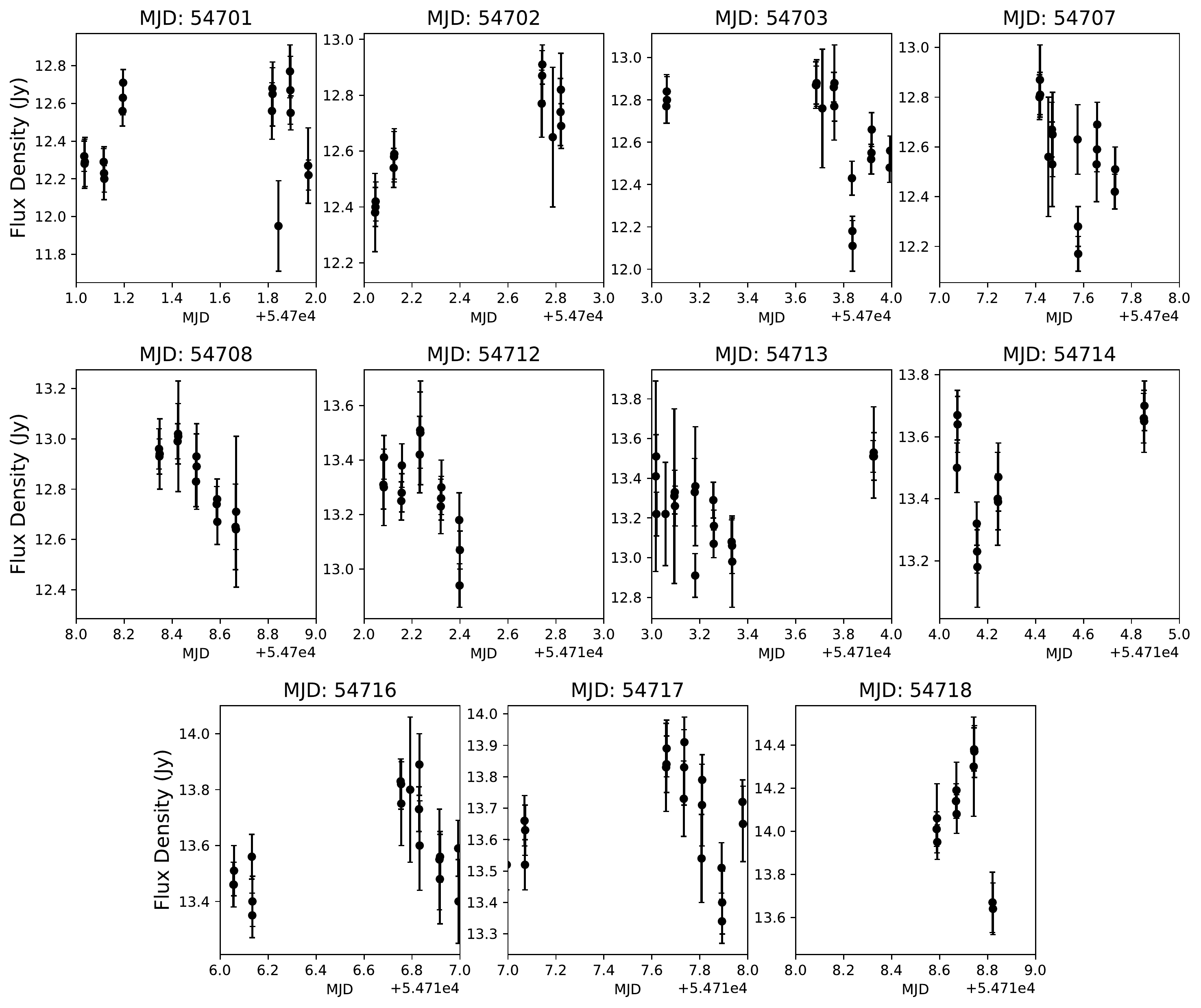}
  \caption{\label{f8}
    15 GHz lightcurves on days when significant variability was observed. $t_d$ for MJD 54708 is larger than IDV timescales.}
\end{figure*}

Long-term flux variations can be caused by a combination of mechanisms arising from intrinsic and extrinsic factors.  Intrinsic mechanisms are mechanisms that are inherent to the blazar jets, like blobs of plasma moving through the helical jet magnetic field \citep{Marscheretal2008}, giving rise to variable compression and polarization \citep[]{Marscheretal2008, Raiterietal2013, Gaur2014} or shocks in the helical jets \citep{Larionovetal2013}. Extrinsic factors include the geometrical effects associated with the change in our viewing angle to a moving emission region. This causes variable Doppler boosting of the emitted radiation \citep[]{Villataetal2009, Larionovetal2010, Raiterietal2013} which is then observed as long-term variations. Both the mechanisms can affect the long-term variability of the light curves \citep{Pollacketal2016} and are often difficult to distinguish.

Significant variability on the multi-year scale is found in all three radio wavebands. In each of the four radio segments: Flare 1, Flare 2, Quiescent and Plateau, the source was also found to be significantly variable. Generally, the percentage variability amplitude for radio segments (given in Table \ref{t3}) is smaller than that for their optical counterparts. This can be understood as the rise time of the radio fluxes is much slower than that for the optical ones. One reason for this could be the  difference in the emission region sizes of the two wavebands, with the radio emission arising from larger regions \citep[]{1985ApJ...298..114M, 2014ApJ...780...87M}.

We observe from Table \ref{t3} that in all segments but the last, $V_F$ increases with increasing radio frequency and that the timescale of variability decreases with increase of frequency in all the segments. The changes in variability can be due to intrinsic factors like synchrotron cooling and adiabatic expansions of shocks \citep{1985ApJ...298..114M} or fully extrinsic factors such as  interstellar scintillation (ISS) \citep{1995ARA&A..33..163W}. The way in which $V_F$ changes with frequency indicates which of these two processes dominates \citep{2012MNRAS.425.1357G}. In the first three segments, the dependence of $V_F$ on the frequency and the significant correlation between radio and optical band (shown in Section \ref{sec:9}) indicates intrinsic causes for the variations. In the last segment, $V_F$ decreases with increase in frequency; this trend, along with the absence of a significant correlation between optical and radio bands, suggests that ISS could cause much of the variability.

There were some observations made in the 15 GHz band where multiple data-points were taken in one day, thereby allowing us to study IDV. Of the 27 days where more than 10 observations were made in a day, we found IDV in 11 of them. The IDV and timescales in the 15 GHz band is tabulated in Table \ref{t8}. Figure \ref{f8} shows the light curves for the days when IDV was observed. The very short flux-doubling timescales during intra-day measurements suggest that ISS could cause IDV in the 15 GHz band \citep{2002PASA...19...55B}. The influence of ISS on the 15 GHz light curve was observed in a large sample of blazars by \cite{2019arXiv190901566K} which reinforces the idea that the 15 GHz IDVs can be due to ISS. Though simultaneous multiple daily measurements in other radio bands are necessary to conclusively rule out intrinsic effects.

\subsection{Spectral variations}
\label{sec:8}

\begin{figure*}[!ht]
        \subfloat{\includegraphics[width=0.5\linewidth]{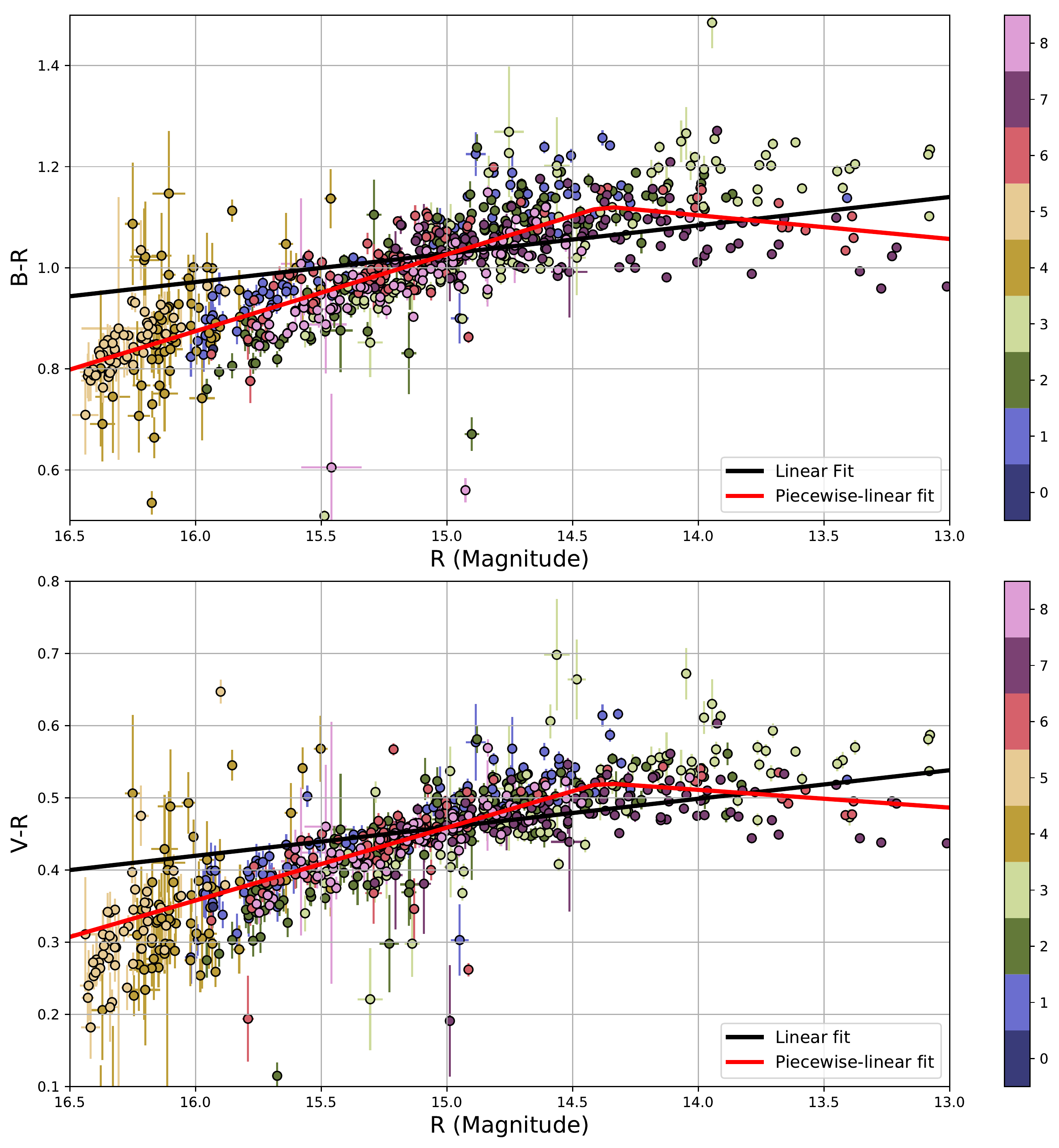}}
                \label{fig:ci1}
        \subfloat{\includegraphics[width=0.5\linewidth]{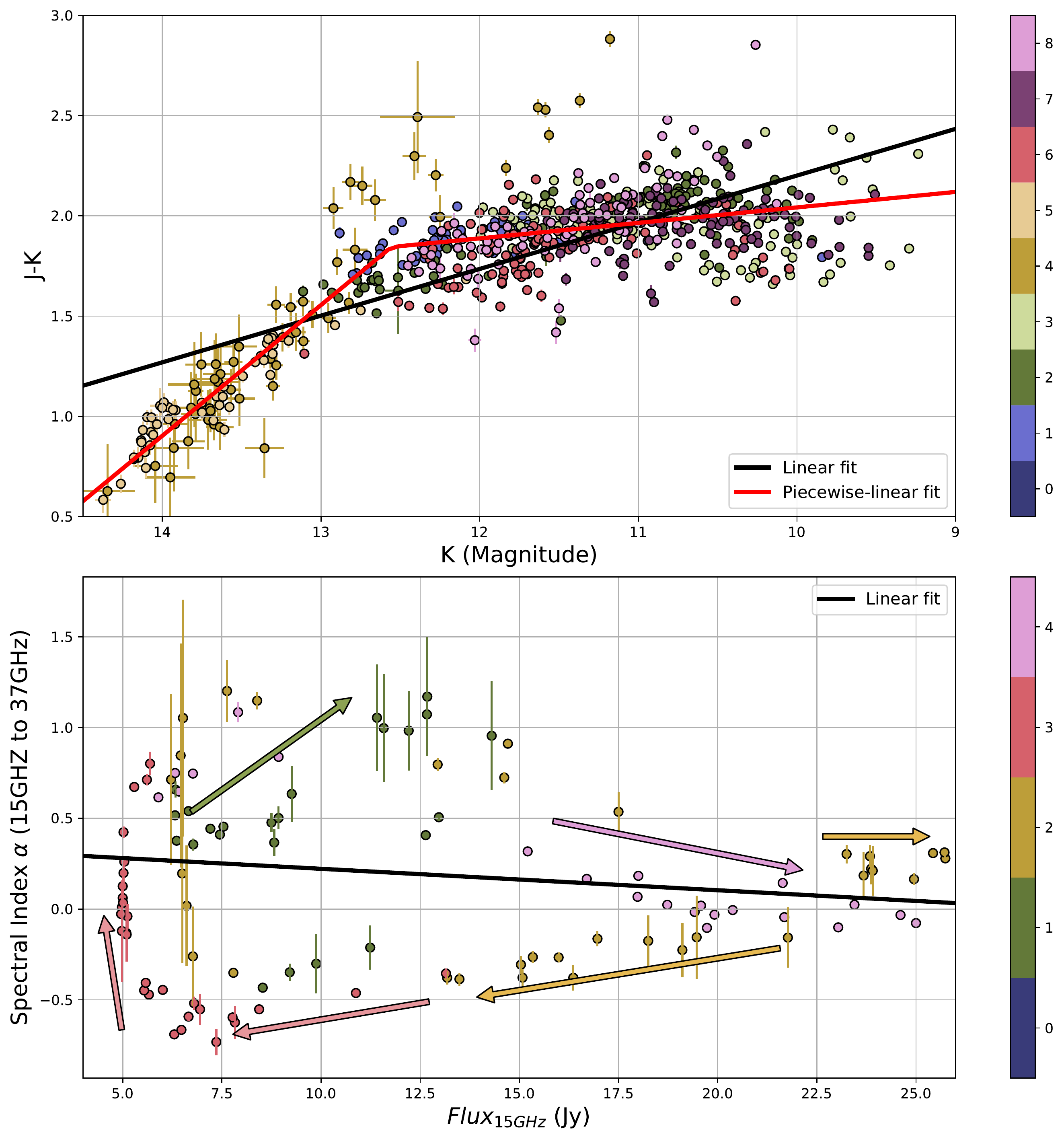}}
                \label{fig:ci2}
        \caption{\textbf{(top-left)} Color-index (B-R) vs R magnitude. \textbf{(bottom-left)} Color-index (V-R) vs R magnitude. \textbf{(top-right)} Color-index (J-K) vs K magnitude. \textbf{(bottom-right)} Radio spectral-index vs 15 GHz flux. The colors indicate the segment in which the data belong and the black lines are the linear fits and the red lines are the piecewise linear fits. In the bottom-right plot, the arrow's slope gives the slope of the best fit line and the arrowhead provides the direction of time.}\label{f2}
\end{figure*}

\begin{deluxetable*}{cccccc}[!ht]
\tablecaption{\label{t4} Fit-parameters for the color-index vs magnitude plots.}
\tablecolumns{6}
\tablenum{5}
\tablewidth{0pt}
\tablehead{
  \colhead{Color Index vs Band} & \colhead{Model} & \colhead{Parameters} & \colhead{$\chi^2$} & \colhead{AIC} & \colhead{BIC}
}
\startdata
B-R vs R (mag) & linear           & $m = -0.056 \pm 0.003$            & 140.69  & 3669.40 & 3678.61\\
               & piecewise-linear & $m_1 = 0.046 \pm 0.008$           & 95.68   & 3389.75 & 3412.79\\
               &                  & $m_2 = -0.15 \pm 0.007$           &         &         &        \\
               &                  & break = $14.37 \pm 0.04$ (mag)    &         &         &        \\
\hline
V-R vs R (mag) & linear           & $m = -0.039 \pm 0.002$            & 51.79   & 2865.77 & 2874.94\\
               & piecewise-linear & $m_1 = 0.025 \pm 0.004$           & 33.60   & 2558.08 & 2581.01\\
               &                  & $m_2 = -0.101 \pm 0.004$          &         &         &        \\
               &                  & break = $14.39 \pm 0.03$ (mag)    &         &         &        \\
\hline
J-K vs K (mag) & linear           & $m = -0.232 \pm 0.007$            & 1616.25 & 4990.81 & 4999.84\\
               & piecewise-linear & $m1 = -0.077 \pm 0.009$           & 1339.96 & 4870.27 & 4892.84\\
               &                  & $m2 = -0.65  \pm  0.03$           &         &         &        \\
               &                  & break = $12.55 \pm 0.06$ (mag)    &         &         &        \\
\hline
\enddata
\end{deluxetable*}

We calculated the (B--R), (V--R) and (J--K) color indices of the dataset to examine the color variability. The color-index--magnitude diagrams are given in Figure \ref{f2}. It is clearly seen that all the color indices decrease with increase in the magnitude, illustrating a redder-when-brighter trend, though they show modest saturations at lower magnitudes. To check whether the saturation is significant, we fitted the data with linear and piecewise-linear functions and calculated the Akaike Information Criterion (AIC) (\citealt{1100705}) and Bayesian Information Criterion (BIC) (\citealt{schwarz1978}) of the models to determine which one explains the data best. The AIC and BIC are given by:

\begin{equation}
  AIC = -2~ln(\mathcal{L}) + 2k,
\end{equation}
\begin{equation}
  BIC = -2~ln(\mathcal{L}) + k~ln(N),
\end{equation}

where $\mathcal{L}$ is the likelihood function of the data, $k$ is the number of parameters in the model and $N$ is the number of data points. Models with lower AIC and BIC are preferred. The model parameters and the values of AIC and BIC are given in Table \ref{t4}. We see that in all the cases the piecewise linear model describes the data better. We also observe the color index to decrease (weakly) as the magnitude decreases at a low magnitude in the (B--R) vs R and (V--R) vs R plots (bluer-when-brighter trend). This negative slope is contributed by points from the flare in segment 7 (May 2014). A linear fit of the high flux points ($<14.5$ mag) in this segment gives the slope of (B--R) vs R equal to $0.0729 \pm 0.0008$. Interestingly, we do not see this trend in the segment 3 flare, where, even though we see saturation, the trend remains nominally redder-when-brighter with a slope of $-0.0226 \pm 0.0005$. The negative slope is also not seen in the (J--K) vs K plot. Earlier studies also showed that the redder-when-brighter trend is more commonly observed in the faint state of the blazar \citep[]{Raiterietal2008, Zhouetal2015}. Previously, a bluer-when-brighter trend was observed in the (B--R)vs R plot during a 2007 flare \citep{Raiterietal2008, Zhaietal2011}.

The color-index vs magnitude plots can indicate the interplay between the jet emission and the accretion disk emission as the thermal emission of the accretion disk is inherently bluer than the jet emission \citep{2013EPJWC..6105001G}. Thus, a redder-when-brighter trend implies that the jet emission dominates the disk emission. Once the jet emission totally outshines the disk emission, a further increase in the jet brightness normally results in a decrease in the color index with an increase in flux, which is due to particle acceleration. When this jet flaring phase passes, the accelerated particles lose energy by radiative cooling which follows the same path in the color-magnitude diagram like the one due to particle acceleration. These two processes combined can show a bluer-when-brighter trend \citep{2017ApJ...844..107I}. Another explanation for bluer-when-brighter trend in bright state of the blazar was given by \citet{2007A&A...470..857P} where the authors attribute it to an increase in Doppler factor that blue-shifts the spectrum. We observed the jet emission completely swamp the disk emission at a magnitude of $\sim$14.3 along with signs of particle acceleration and radiative cooling during 2014 flare.

For the radio bands, we computed the spectral index by fitting the co-temporal flux densities of different radio bands to a power law model $F_{\nu} \propto \nu^\alpha$, where $F_{\nu}$ is the flux density at a frequency $\nu$ and $\alpha$ is the spectral index. This spectral index vs flux-density plot is also given in Figure \ref{f2} (bottom-right). We see the spectral index increasing with 15 GHz flux-density during Flare 1 with the index varying from 0.5 to 1. During flare 2, this index mostly remained constant with time. During the quiescent state, we see the index to be constant ($< 0$) at higher flux density values. At lower flux values, the index varies between $-0.5$ to $0.5$. In the plateau region, the radio spectral index value is close to 0 and does not change much with the flux density. The best fit line for the entire stretch is almost horizontal ($m=-0.011\pm 0.007$).

\subsection{Correlation Measurements}
\label{sec:9}

\begin{figure*}[!]
  \subfloat{\includegraphics[width=\linewidth]{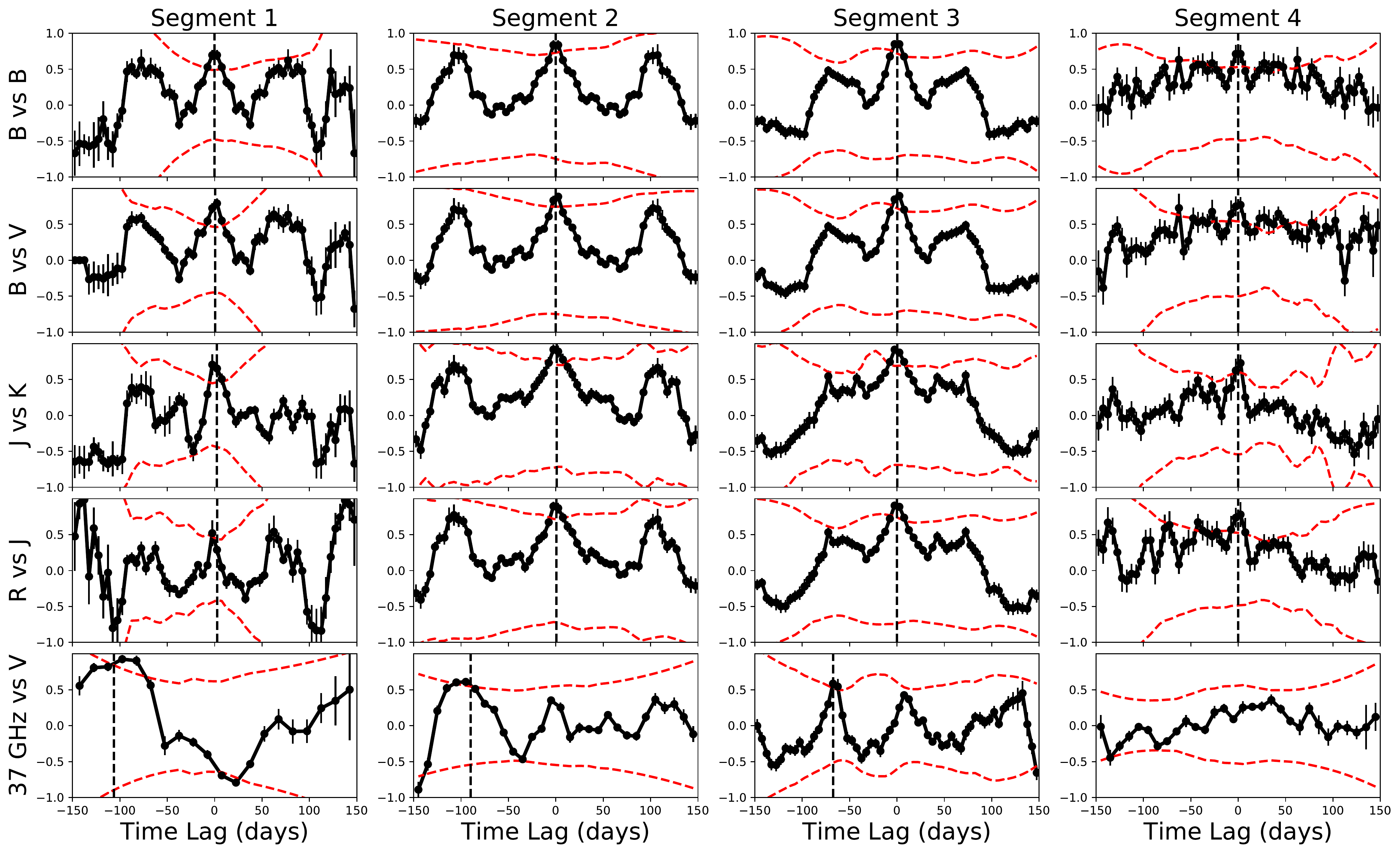}}
  \label{fig:corr1}
  \qquad
  \subfloat{\includegraphics[width=\linewidth]{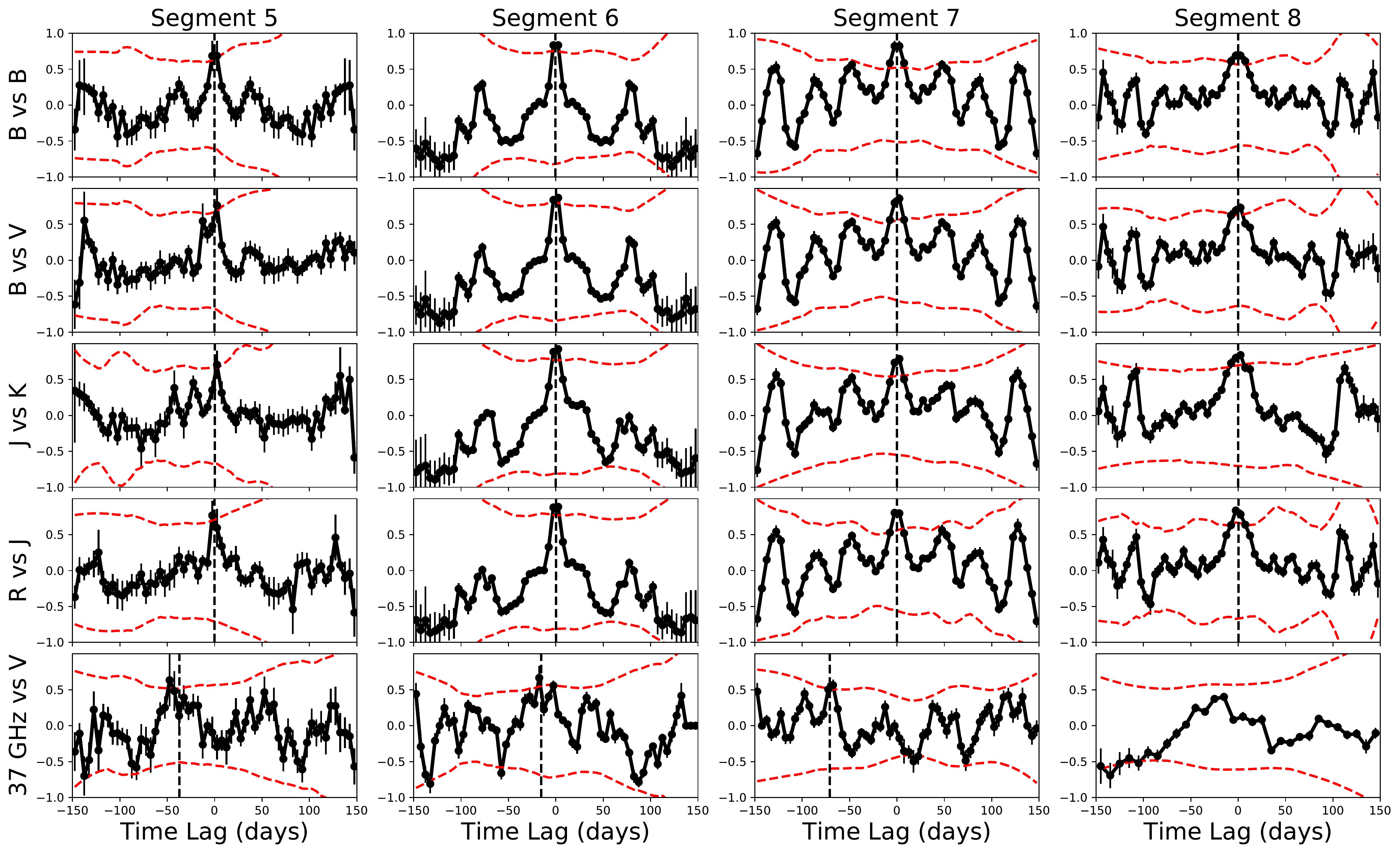}}
  \label{fig:corr2}
  \caption{DCF between different wavebands in different temporal segments. The dashed red lines show the 99.9\% confidence level and the dashed vertical line is the most significant maximum of the DCF.  A DCF peaking at negative lag means the first light curve is lagging behind the second. }\label{f3}
\end{figure*}

\begin{figure*}[htb]
    \includegraphics[width=\linewidth]{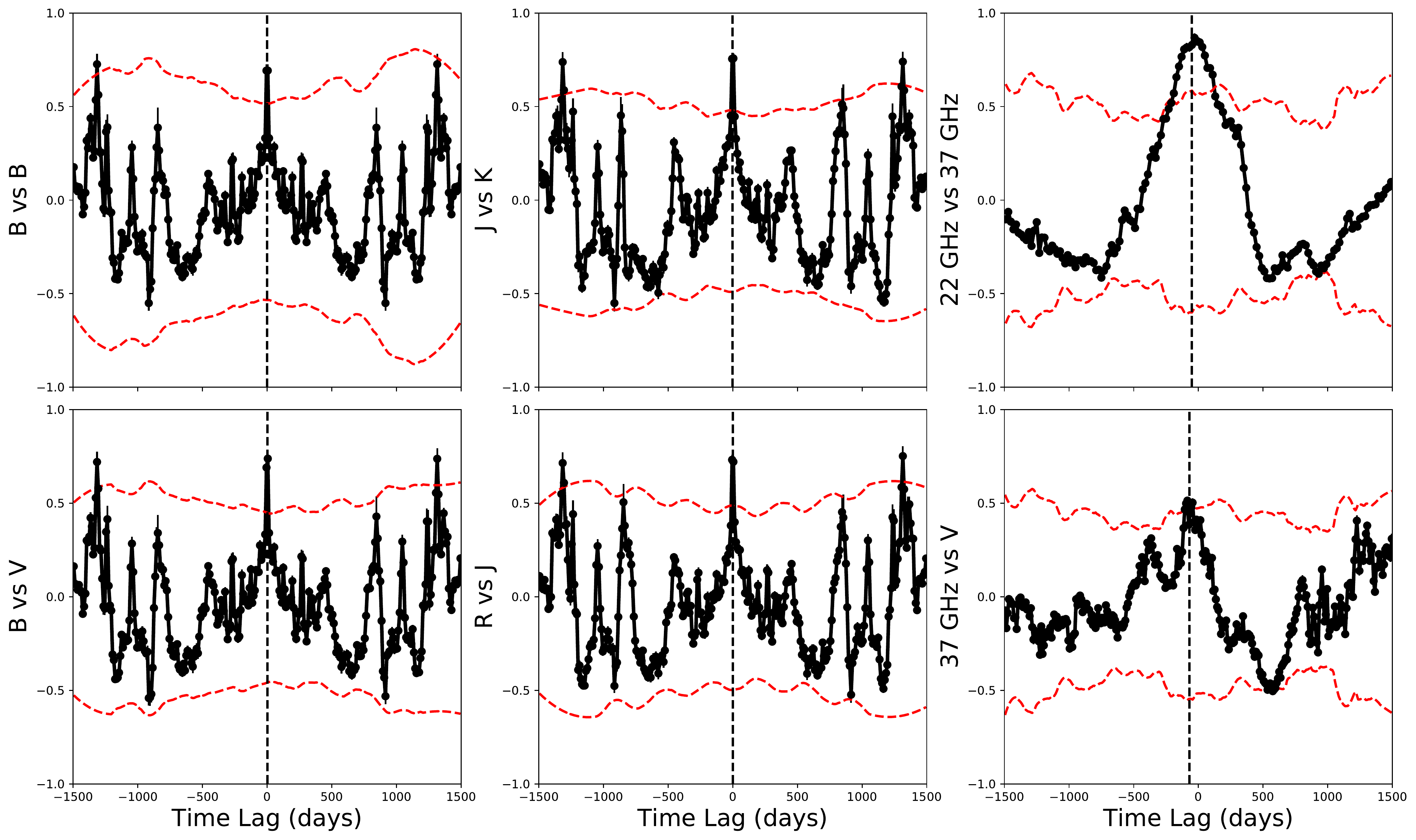}
    \caption{\label{fig:corr3} As in Fig. \ref{f3} for  the totality of the observations.}
\end{figure*}
\begin{figure*}[htb]
    \includegraphics[width=\linewidth]{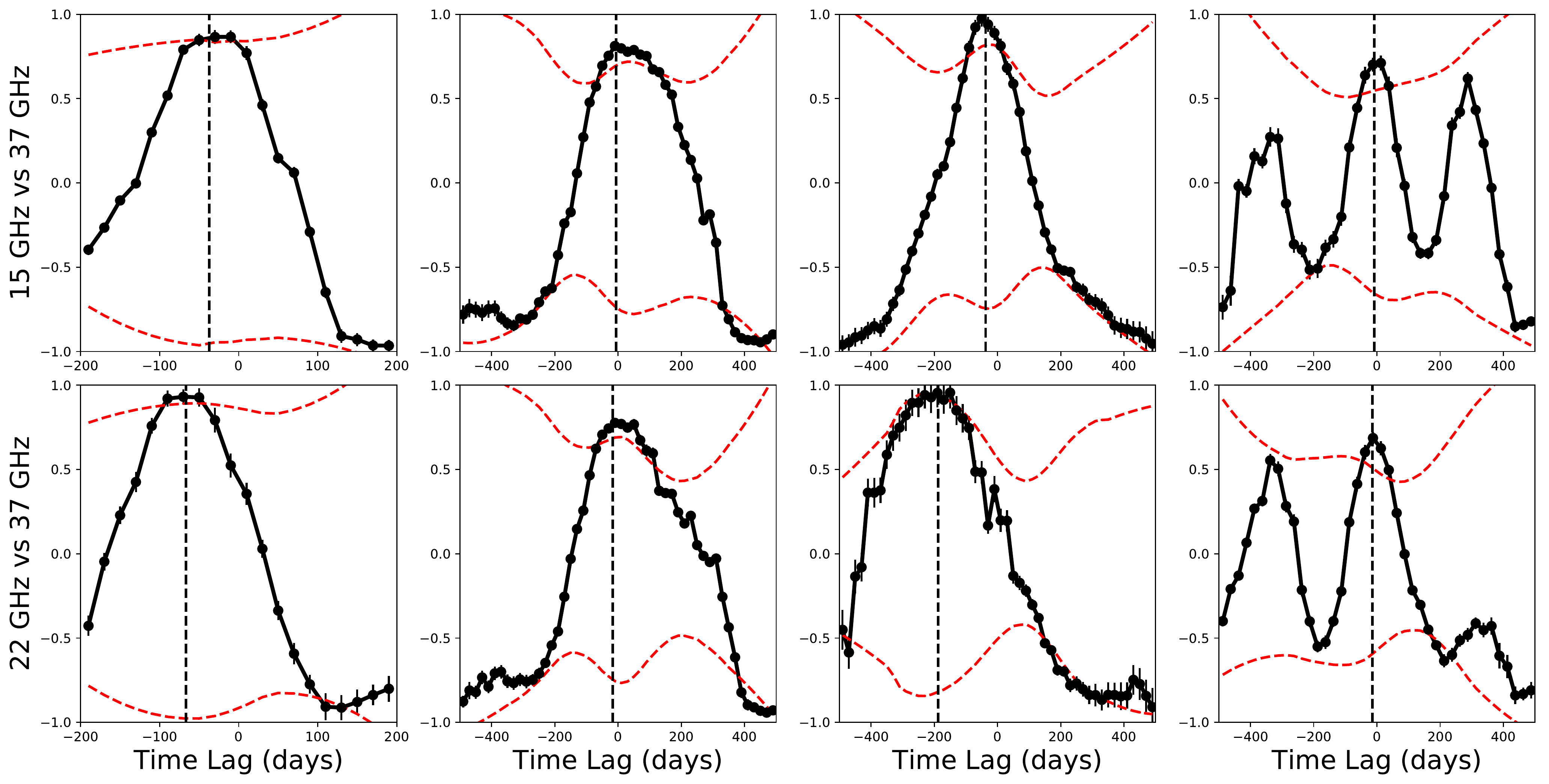}
    \caption{\label{fig:corr4} As in Fig. \ref{f3} for radio bands in radio segments.}
\end{figure*}

We performed correlation measurements on the available time series to search for possible variability timescales and lags between light curves of different wavebands. Since the light curves consist of discrete points, we used discrete correlation functions (DCFs) to analyze the light curves. We used the following definition of the unbinned DCF (UDCF) between the $i^{th}$ datapoint in one band and the $j^{th}$ datapoint in another \citep{1988ApJ...333..646E}:

\begin{equation}
  \label{eq4}
  UDCF_{ij}=\frac{(a_i-\bar{a})(b_j-\bar{b})}{\sqrt{\sigma_a^2 \sigma_b^2}},
\end{equation}

where $a_i$ and $b_j$ are points on the first ($a$) and second ($b$) light curve, respectively, while $\bar{a}$ and $\bar{b}$ are the averages and $\sigma_a$ and $\sigma_b$ are the standard deviations of the flux values from two light curves. Next, we bin the correlation function by averaging the time delay $\Delta t_{ij}=t_{bj}-t_{ai}$ lying in the range $\tau-\frac{\Delta \tau}{2}\leq \Delta t_{ij}\leq\tau+\frac{\Delta \tau}{2}$ ($\tau$ is the time lag and $\Delta \tau$ is the bin width) and evaluate the DCF as:

\begin{equation}
  DCF(\tau)=\frac{1}{n}\sum UDCF_{ij}(\tau).
\end{equation}

Since the emission from a blazar is a non-stationary statistical process, its mean and variance also change depending on the length of the light curves being used. Following \cite{WhitePeterson1994}, the means ($\bar{a}$ and $\bar{b}$) and variances ($\sigma_a$ and $\sigma_b$) in Eq. \ref{eq4} were calculated using only the points that fall within a given time-lag bin. The error in each bin is the standard deviation of the number of points in the bin that were used to determine the DCF and is given by:

\begin{equation}
  \sigma_{DCF}(\tau)=\frac{1}{M-1}\sqrt{\sum_{k=1}^M(UDCF_k-DCF(\tau))^2}.
\end{equation}

The DCFs in the eight individual segments are shown in Figure \ref{f3} and the DCFs for the full span of data (3000 days) are shown in Figure \ref{fig:corr3} for a few selected waveband pairs. A linear baseline was substracted from each of the light-curves (de-trending) while calculating the DCFs \citep{Welsh1999}. Table \ref{t5} shows the time lags where the DCFs peak. The significance of a DCF peak was calculated using the method in \cite{2014MNRAS.445..437M}. A thousand lightcurves were generated in each waveband following the power spectral density (PSD) and the flux distribution function (PDF) of the original light curve \citep{2013MNRAS.433..907E}. From the distribution of the DCF between the simulated and the original light curve, at each lag, the threshold for 99.9\% significance was estimated. Such significant peaks were fitted with Gaussian functions to obtain the peak positions and their uncertainties. We observe wide variations in the shape of the correlation functions in various segments, ranging from good correlations at zero time lag to relatively flat DCFs and also DCFs peaking at non-zero lags.

\begin{deluxetable*}{cccccc}[!ht]
\tablecaption{\label{t5} Time lag (in days) at DCF peak for the different wavebands\tablenotemark{1}}
\tablecolumns{6}
\tablenum{6}
\tablewidth{0pt}
\tablehead{
  \colhead{Seg} & \colhead{B vs B} & \colhead{B vs V} & \colhead{J vs K} & \colhead{R vs J}  & \colhead{37 GHz vs V \tablenotemark{2}}
}
\startdata
      Tot     & $1.0 \pm 2.6$  (0.83) & $3.1 \pm 2.3$  (0.86) & $-1.6 \pm 3.2$ (0.88) & $0.0 \pm 2.9$  (0.86) &   $-103.4 \pm 7.9$ (0.52) \\
      1       & $0.0 \pm 0.7$  (0.69) & $0.5 \pm 0.9$  (0.79) & $2.2 \pm 1.3$  (0.70) & $2.6 \pm 2.3$  (0.89) &   $-106.2 \pm 3.6$ (0.92) \\      
      2       & $0.0 \pm 0.3$  (0.84) & $-0.2 \pm 0.8$ (0.88) & $1.0 \pm 0.8$  (0.92) & $0.6 \pm 0.8$  (0.89) &   $-89.7 \pm 1.9$  (0.61) \\      
      3       & $0.3 \pm 0.3$  (0.85) & $0.4 \pm 0.2$  (0.90) & $0.4 \pm 0.3$  (0.92) & $0.0 \pm 0.4$  (0.91) &   $-67.3 \pm 0.7$  (0.58) \\
      4       & $0.0 \pm 0.1$  (0.71) & $0.0 \pm 0.0$  (0.77) & $-0.1 \pm 0.2$ (0.73) & $0.0 \pm 0.1$  (0.78) & 	 -                       \\
      5       & $0.0 \pm 0.4$  (0.68) & $0.0 \pm 0.4$  (0.76) & $0.1 \pm 0.4$  (0.71) & $-0.2 \pm 0.3$ (0.77) &   $-37.2 \pm 2.29$ (0.63) \\
      6       & $-0.3 \pm 1.6$ (0.83) & $0.1 \pm 0.9$  (0.87) & $0.0 \pm 1.1$  (0.92) & $0.4 \pm 0.4$  (0.89) &   $-15.4 \pm 1.1$  (0.79) \\
      7       & $0.0 \pm 0.1$  (0.82) & $0.0 \pm 0.1$  (0.86) & $-0.2 \pm 0.4$ (0.79) & $0.1 \pm 0.4$  (0.80) &   $-73.9 \pm 1.0$  (0.48) \\
      8       & $0.0 \pm 0.6$  (0.69) & $0.0 \pm 0.7$  (0.73) & $-0.1 \pm 0.4$ (0.84) & $0.2 \pm 0.6$  (0.84) &   -                       \\
\enddata
\tablenotetext{1}{The correlation value at the peak is given in the bracket.}
\tablenotetext{2}{Not all 37 GHz vs V band correlations are significant.}
\end{deluxetable*}

\begin{deluxetable}{ccc}[t]
\tablecaption{\label{t11} Time lag in days at DCF peak for different radio bands.}
\tablecolumns{3}
\tablenum{7}
\tablewidth{0pt}
\tablehead{
\colhead{Period} & \colhead{15 GHz vs 37 GHz} & \colhead{22 GHz vs 37 GHz}
}
\startdata
Flare 1   & $-37.1 \pm 2.6$ (0.93)  & $-66.9 \pm 0.9$  (0.98)  \\
Flare 2   & $-6.6 \pm 1.3$  (0.83)  & $-16.7 \pm 1.0$  (0.80)  \\
Quiescent & $-37.6 \pm 2.0$ (0.97)  & $-187.2 \pm 3.8$ (0.96)  \\
Plateau   & $-8.2 \pm 1.7$  (0.74)  & $-14.6 \pm 1.9$  (0.69)  \\
\enddata
\end{deluxetable}

We observe significant correlations at zero lag for optical/IR bands. All possible optical--optical DCFs show a similar structure, with the DCF peaking at zero lag and then rapidly falling as the lag increases. IR--IR and optical--IR DCFs follow their optical counterparts with the DCFs peaking at zero-lag implying that the variations in the optical/IR bands are dominated by emission from a single region in the blazar. Similar structure was also observed in the optical/IR DCFs considering the full 3000 days stretch (Figure \ref{fig:corr3}). 

The autocorrelation function for the radio bands revealed no characteristic timescales of variability. The 22--37 GHz DCFs show a clear peak at lags of the order of 15--190 days with the 37 GHz emission leading the 22 GHz emission. This lag is visually evident during the Flare 1 period in Figure \ref{f1}. The 15 GHz band also lags behind the 37 GHz band by 5--40 days. The DCF of 15 and 22 GHz vs 37 GHz band is given in Figure \ref{fig:corr4} and the position of the peak of the DCF is given in Table \ref{t11}. The 22--37 GHz lag is consistently higher than the 15--37 GHz lags which cannot be explained by standard shock-in-jet model. However, the lags involving the 15 GHz data are not as well determined because of the sparseness of data, particularly at the critical epochs close to the flare peaks.

The inter-band cross-correlation function between radio and optical bands (37 GHz vs V) shows the DCFs peaking at a lag of $\sim$ 100 days, with the optical emission leading the radio emission. From Table \ref{t5}, we see that the peak of the DCFs varies significantly from segment to segment, from as low as 15 days (in Segment 6) to as high as 106 days (in Segment 1).

\section{Discussion}
\label{sec:10}

\begin{figure*}[htb]
        \subfloat{\includegraphics[width=0.5\linewidth]{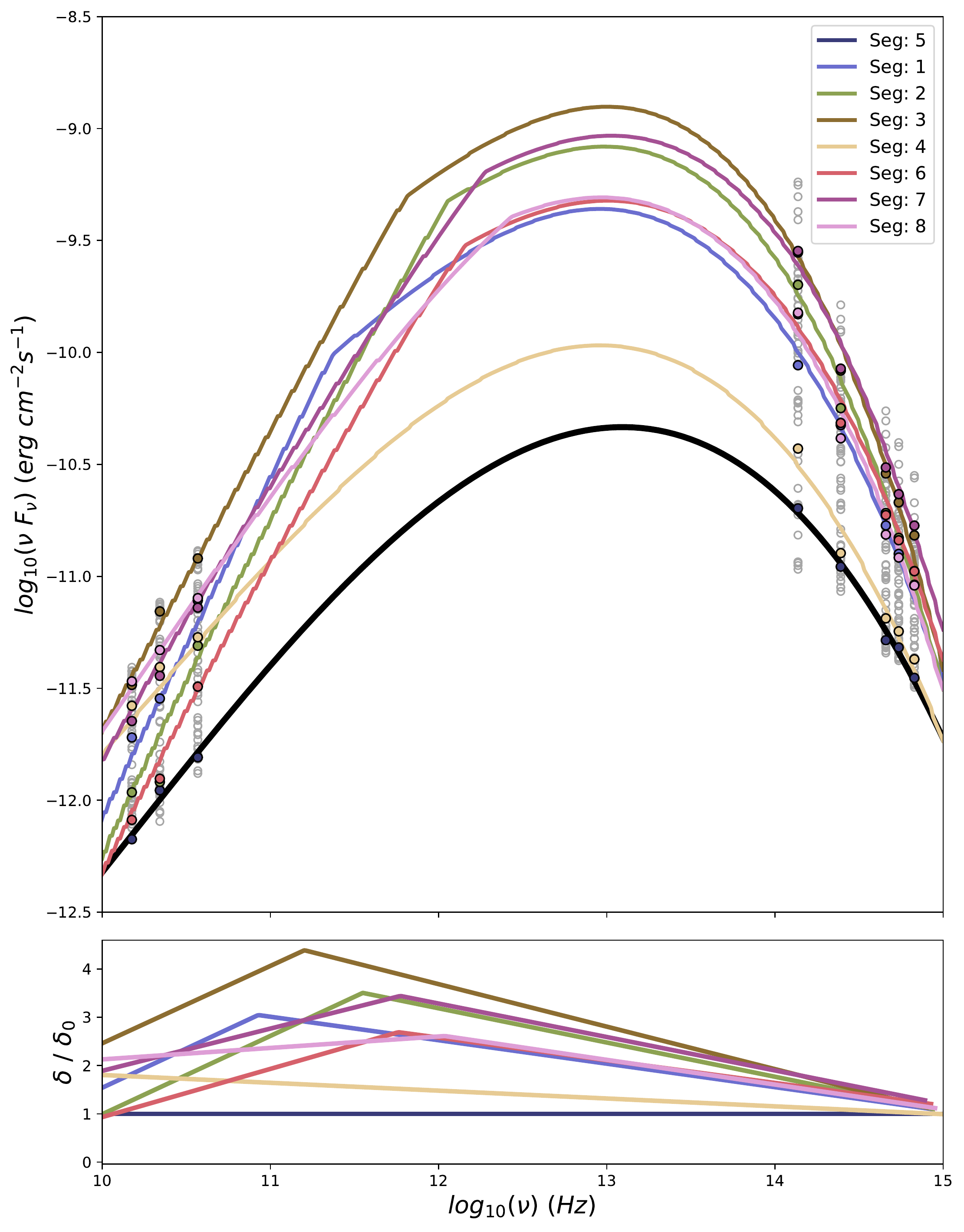}}
                \label{fig:df1}
        \subfloat{\includegraphics[width=0.5\linewidth]{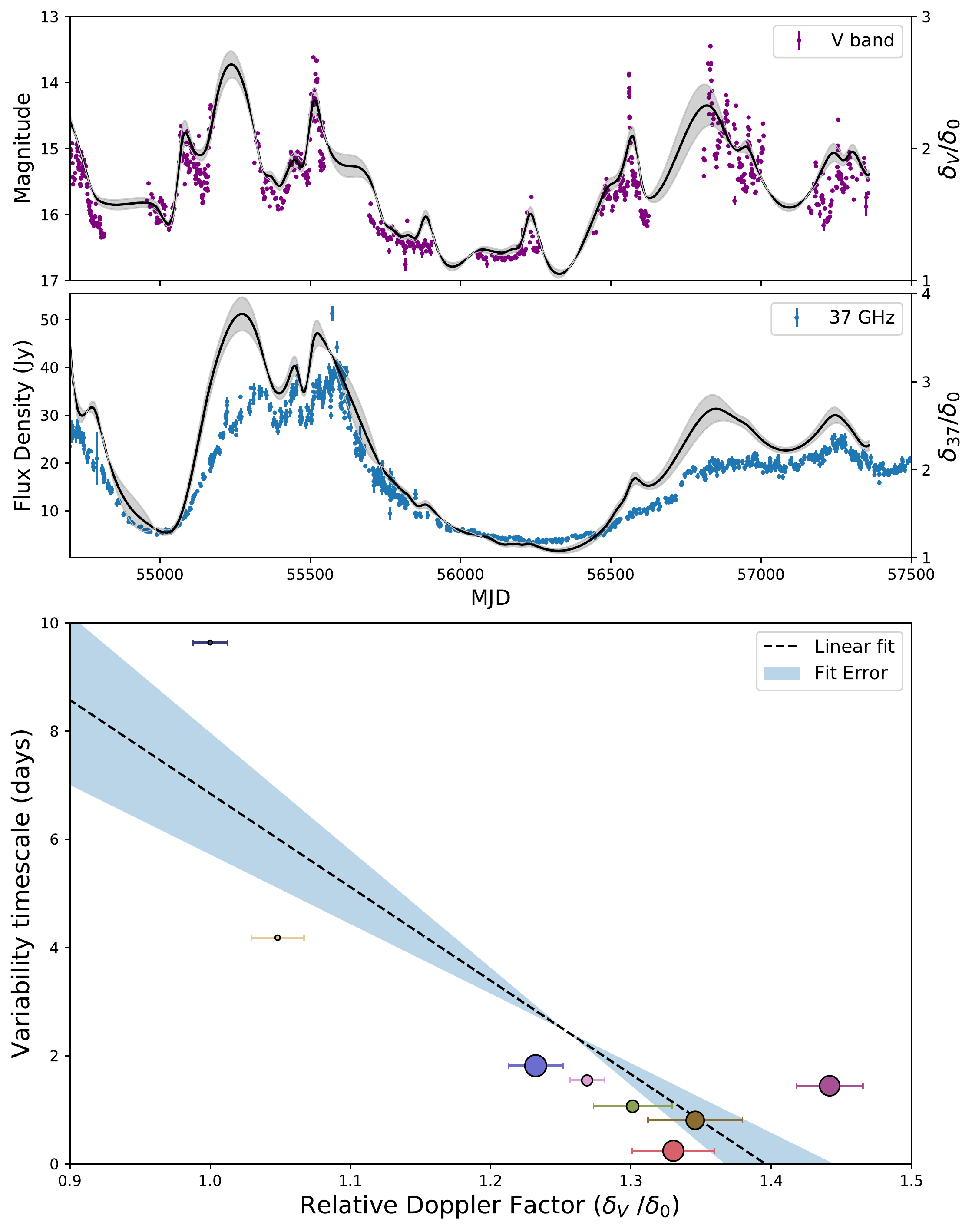}}
                \label{fig:df2}
        \caption{\textbf{(top-left)} Spectral model, with the baseline (approximated by average Segment 5) emission in thick black; gray points represents all the observations and colored points represent the mean value for each segment. The colored lines are spectra obtained by Doppler boosting the baseline fluxes by a frequency dependent piecewise-linear Doppler factor given in the \textbf{(bottom-left)} panel. Temporal variations in Doppler factor in \textbf{V} band \textbf{(top-right)} and 37 GHz band \textbf{(middle-right)}. The gray region represents $1\sigma$ estimated error on the Doppler factor. \textbf{(bottom-right)}. Variability timescale vs Doppler factor for V band where larger circles correspond to higher variability amplitudes. Color scheme follows from before.}\label{f7}
\end{figure*}

We observe clear signatures of jet emission dominating the disk emissions. The unenhanced disk emission are more stable than the boosted jet emission, significantly higher variability in the IR bands as compared to optical bands could be a signature for a disk emission contribution, which was not observed. A bluer-when-brighter trend in the color-index vs magnitude diagram was not observed in the low flux state, reinforcing the idea that jet emissions dominate even in the low flux state. This claim is further substantiated by observation of significant correlations at zero lag between optical and IR bands in the low flux state (Segment 4, Segment 5). These strong correlations indicate that the emission regions are co-spatial even in the low flux state and can be used to argue against a significant contributions from the accretion disk. In both optical and IR wavebands, the index decreases with magnitude, indicating a redder-when-brighter trend which has been seen to be a general trend for this source at brighter levels \citep{Villataetal2006}. Bluer-when-brighter trends and saturation of color indices were observed at the high flux level in Seg. 7, which can be signatures of particle acceleration and radiative cooling.

The source was observed to be variable across different timescales. The long-term variations are likely to be caused by a combination of what can be considered as extrinsic factors (e.g., changes in Doppler factor due to changes in viewing angles) superimposed on completely intrinsic factors (e.g., motion of denser plasma through an enhanced magnetic field or adiabatic expansion of shocks). The extremely short variability timescales for some radio IDV observations suggest that these could arise, from interstellar scintillation. Variability due to ISS is mainly observed in lower frequency wavebands even though for some sources 15 GHz variability due to ISS has been observed \citep[e.g.,][for Ton 599]{2008A&A...489L..33S}. From the variability timescale, one can constrain the emission region size using causality arguments. The upper-limit on the emission region size is given by:

\begin{equation}
  \label{eq5}
  R_{max}=\frac{c\delta\tau_{var}}{1+z},
\end{equation}

where $\delta$ is the Doppler factor and $z$ is the redshift. We adopted $\delta \approx 30$ \citep{2009A&A...494..527H} in our calculations. Estimation of the linear size gives the maximum angle subtended at the observer using $\Phi_{max}= R_{max}(1+z)^2/r_{bol}$, where the luminosity distance, $r_{bol}$, was calculated using Hubble's constant $H_0=67.3$ km s$^{-1}$ Mpc$^{-1}$, the total mass fraction, $\Omega_M=0.315$, and a flat cosmology \citep{PlankCollab2014}, which gives $r_{bol}=5.65$ Gpc. The upper-limits on the linear size and the angle subtended at the observer are given in Table \ref{t7}. Calculations for the optical regions are based on V band data whereas in the case of radio the data from the 37 GHz band was used because it is very densely sampled, thus allowing us to detect the shortest possible flux doubling time.

The radio vs optical DCF peaks at non-zero lags ($\tau_{delay}$) with the optical emissions leading the radio emissions, implying that the radio and optical emission regions are not co-spatial with the optical/IR emission region being closer to the base of the jet. This is consistent with the observation that optical emission regions are smaller than radio emission regions (Table \ref{t7}). These observations are in accordance with standard shock-in-jet models where higher frequencies are emitted closer to the shock front while lower frequencies are produced from larger volumes that extend further away from the shock \citep[e.g.,][]{1985ApJ...298..114M, Marscheretal2008}. This could also be understood as a manifestation of the position offset of optically thick features that can be interpreted as a frequency dependent shift of the self-absorbed core of the jet \citep[e.g.,][]{1998A&A...330...79L, 2012A&A...545A.113P}. The linear seperation of the V and 37 GHz emission region can be estimated using the relation \citep{2010ApJ...722L...7P, 2017MNRAS.468.4478L}:

\begin{equation}
  D^{opt-radio}_{max} = \frac{\beta_{\mathrm{app}} c\ \tau_{delay}}{\sin \theta(1+z)}
\end{equation}

where $\beta_{\mathrm{app}}$ is the apparant jet speed and $\theta$ is the viewing angle. Using a range of 15--100 days for $\tau_{delay}$ from Table \ref{t5}, the maximum linear separation between the emission regions ($D^{opt-radio}_{max}$) is estimated to be in the range of 4.06 (for Seg. 6) to 27.04 pc (for Seg. 1) and the corresponding projected seperation varies from 0.18--1.20 pc using a viewing angle of $\theta=1.3^0$. The resulting angular seperation is 0.022--0.151 mas. The simplest jet geometry is that of a conical jet. It cannot however explain the change in the seperation between emission regions. In a conical jet geometry, the distance of an emission region from the central engine can be calculated using $d_{ce} \approx c \Gamma \delta \tau_{var} / (1+z)$ \citep{Abdoetal2011} assuming the emission region fills the cross-section of the jet and the opening angle  $\Phi_{op}\approx 1/\Gamma$. The obtained $d_{ce}$ is given in Table \ref{t7} and the seperation comes out to be 1.4 pc. Thus, the conical jet model also severely underestimates the seperation between the emission regions as it does not take into account the jet collimation. An alternative model is that of an inhomogeneous curved jet, where synchrotron radiation of decreasing frequency is produced in an outer and wider jet region which changes orientation with time. It is possible that the long-term variability behavior of 3C 454.3 during our extended observation is dominated by geometrical effects that also leads to temporal delays between the radio and optical bands.

\begin{deluxetable}{ccc}[htb]
  \tablecaption{\label{t7} Physical parameters of emission regions}
  \tablecolumns{3}
  \tablenum{8}
  \tablewidth{0pt}
  \tablehead{
    \colhead{Parameter} & \colhead{Optical} & \colhead{Radio}\\
    \colhead{} & \colhead{551 nm} & \colhead{37 GHz}
  }
  \startdata
  $t_{var}$ (days)        & 0.6      & 3.2    \\
  $R_{max}$ (pc)          & 0.008    &  0.043 \\
  $\Phi_{max}$ ($\mu as$) & 1.02      & 5.54    \\
  $d_{ce}$ (pc)          & 0.16   & 0.86 \\
  \enddata
\end{deluxetable}

A very basic curved jet model involves the assumption that the Doppler factors of the different emission regions are different. We assume that there is a baseline emission ($\mathcal{F}_{\nu0}$) which has a constant Doppler factor ($\delta_0$) for all the different frequencies. This emission is Doppler boosted by a frequency dependent Doppler factor ($\mathcal{F}_{\nu} \propto \delta_{\nu}^3 \mathcal{F}_{\nu0}$) that we observe. We construct a baseline emission by taking the minimum fluxes ($\nu \mathcal{F}_{\nu}$) for each waveband and fitting them using a log-parabola model. We do not model the thermal emission from the disk since the variability, color-index and correlation analyses all show that even in the low flux state the jet emission dominates that from the accretion disk. Variations between bands are obtained using the relativistic invariance of $\mathcal{F}_{\nu}/\nu^2$ which gives us the corresponding baseline frequency ($\nu_0$) for each observations of $\mathcal{F}_{\nu}$. Then $\delta_{\nu}$ is estimated using the relation $\delta_{\nu} = \delta_0 (\nu/\nu_0)$; this estimated Doppler factor increases with frequency in the radio bands and decreases with frequency in the optical regime. We model it using a piecewise linear function and use this model of Doppler factor to obtain the average spectral energy distribution (SED) in each segment. The obtained SEDs and the Doppler factor model is given in Figure \ref{f7} (left) considering the average Segment 5 emission as a baseline. This analysis for 3C 454.3 follows that of \citet{2017Natur.552..374R} for the source CTA 102. Assuming that the variability is caused by changing Doppler factors, one can trace the temporal evolution of the Doppler factor in different wavebands. The relative Doppler factors ($\delta_{\nu}/\delta_0$) for the V and 37 GHz bands are shown in Figure \ref{f7} (top-right) and (middle-right) respectively. The Doppler factors are correlated at near-zero lag with the observed flux densities in the respective wavebands resulting in $\delta_V$ leading $\delta_{37}$ by $\sim 100$ days for the total data stretch. Due to Doppler boosting, the variability timescales appear shorter in the observer frame ($\Delta t = \Delta t' / \delta$) and the variability amplitude is larger \citep{1995PASP..107..803U}. Plotting the relative Doppler factor in each of the segments vs variability timescales in Figure \ref{f7} (bottom right), we see that the timescale decreases as the Doppler factor increases ($\rho=-0.07\pm 0.01$). Also, points with high $V_F$ are clustered near the region of high $\delta/\delta_0$ which is consistent with the variability arising from changing Doppler boosting of the emission regions.

\begin{figure*}[htb]
  \centering
        \subfloat{\includegraphics[width=0.50\linewidth]{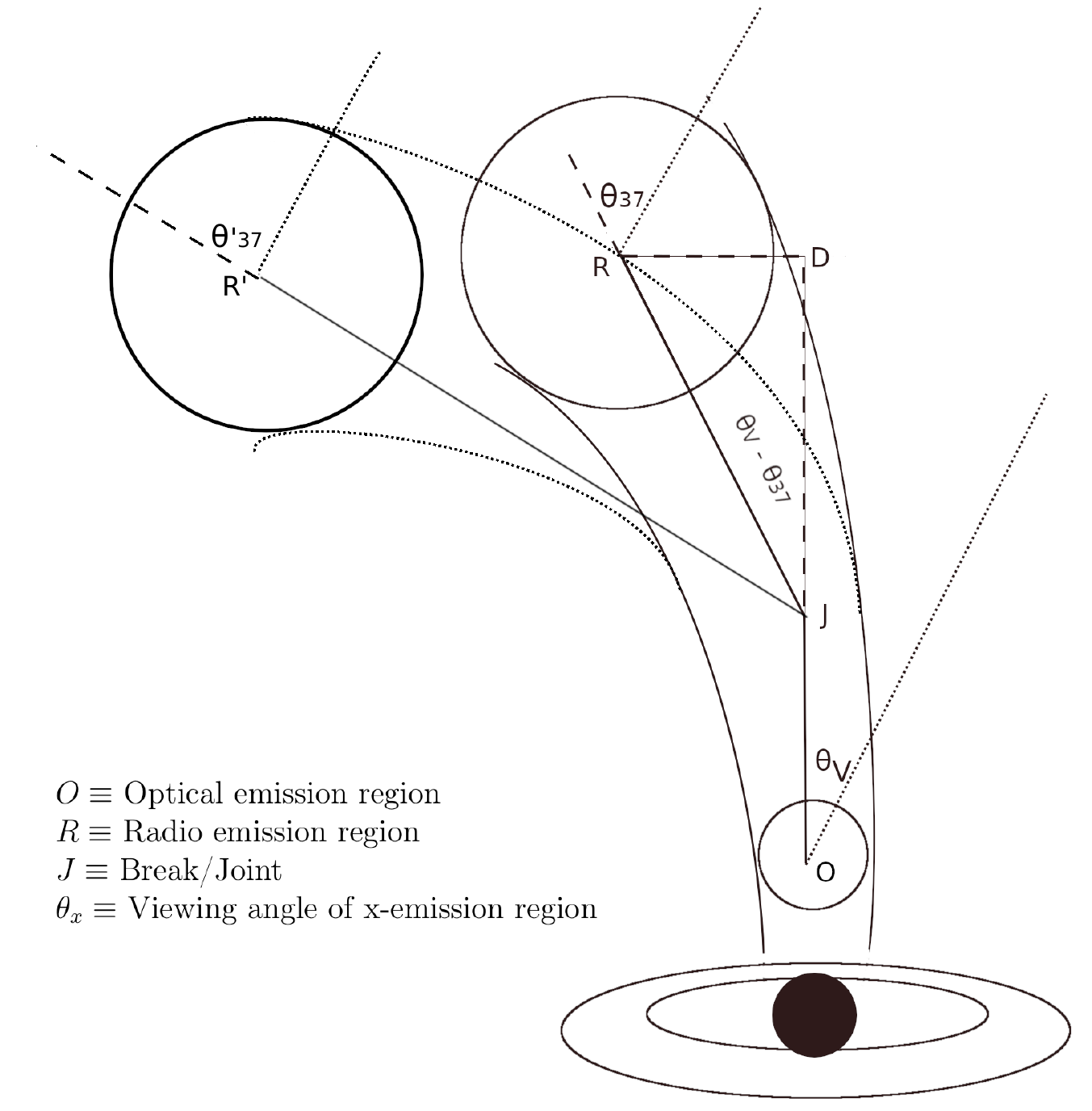}}
                \label{fig:df3}
        \subfloat{\includegraphics[width=0.42\linewidth]{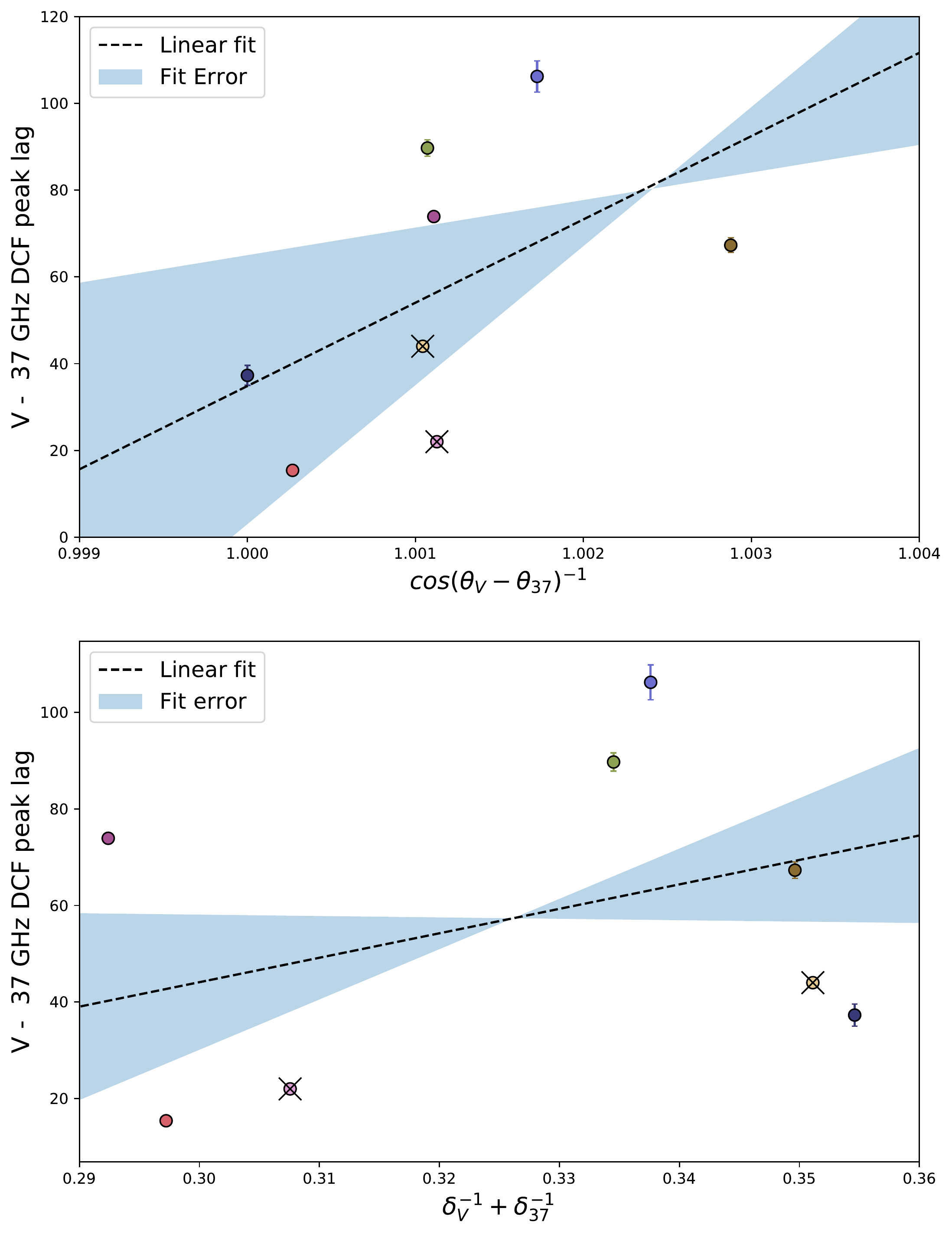}}
                \label{fig:df4}
        \caption{\textbf{(left)} Simple model of a curved jet for 3C 454.3. The dotted line shows the direction of observer. The angles are exaggerated. When the radio emission region is at $R'$, the viewing angle and distance from optical emission region increases (compared to when the emission region is at $R$) thereby increasing the DCF peak lag. \textbf{(right-top)} V vs 37 GHz DCF peak lag vs $cos(\theta_v-\theta_{37})^{-1}$. \textbf{(right-bottom)} V vs 37 GHz peak vs $\delta_V^{-1}+\delta_{37}^{-1}$. The DCF peaks at the crossed points are not $3\sigma$ significant and only significant
          points were used to compute the best-fit line. Color scheme as in Fig. \ref{f2}.}\label{f10}
\end{figure*}

We can construct a rudimentary curved jet model to explain the variation in the radio-optical DCF peak lag (Figure \ref{f10}, left). We assume that the height of the emission region from the central engine remains constant and any disturbance propagates along the length of the jet. We approximate the curved jet as a broken line ($OJR$ in the figure) and assume that the change in $\delta$ to be solely due to the change in viewing angle. In principle, a change in $\Gamma$ can also lead to a change in $\delta$ but it requires very high differential acceleration \citep{2017Natur.552..374R}, and hence we do not consider this option. From geometric considerations, one can see that the distance between optical and radio emission region $OR \propto \frac{1}{cos(\theta_V-\theta_{37})}$ in the rest frame of the jet, where $\theta_x$ is the viewing angle of the emission region $x$. Plotting the lag for the DCF peak between V and 37 GHz against $ \frac{1}{cos(\theta_V-\theta_{37})}$ (Figure \ref{f10}, right-top), we observe a positive correlation ($\rho = 0.5 \pm 0.1$). Due to Doppler boosting, the observed time lag should also be proportional to $\frac{RJ}{\delta_{37}} + \frac{OJ}{\delta_V}$ where assuming $OJ \sim RJ$, we see positive correlation ($\rho=0.36\pm0.37$) between the V--37 GHz DCF peak lag and $\frac{1}{\delta_{37}} + \frac{1}{\delta_V}$ (Figure \ref{f10}, right-bottom). This implies that a changing curvature in the jet could explain the change in DCF peak lags. This analysis required the explicit values of $\delta_0$ and we used $\delta_0 \approx 7$. Using $\Gamma = 20$ \citep{2009A&A...494..527H}, we obtain the maximum viewing angle ($\theta_{max}\approx 6^o$) and the minimum viewing angle ($\theta_{min} \approx 2.3^o$) using $\delta_{\{0,max\}}=[\Gamma(1-\beta\ cos\ \theta_{\{max,min\}})]^{-1}$ where $\delta_{max}$ is the maximum obtained Doppler Factor. The minimum viewing angle comes out to be slightly higher than the values quoted in the literature \citep[$1.3^\circ$ in][]{2009A&A...507L..33P, 2009A&A...494..527H}. The obtained value is not the best esitmate for the viewing angles as there is an inherent ambiguity in the choice of $\delta_0$.  Another possible explanation for the changing radio--optical DCF peak could be due to the motion of standing shocks (localized radio emission regions) in a jet over time due to change in the physical conditions in the jet \citep{2017MNRAS.468.4478L, 2017A&A...597A..80H, 2019MNRAS.485.1822P}.

DCF peaks at lags of 6--180 days are seen between individual radio bands in the present work. One possible explanation for this is the core-shift effect, defined as the apparent systematic outward shift of the VLBI core position with decreasing observation frequency. It does not appear that any simple model can explain why the 15 GHz emission leads the 22 GHz emission in the data that we have collected. While the major flares could be, and probably are, fundamentally produced by shocks propagating down a jet, the variations in both color-indices and temporal gaps between bands require additional complications beyond those provided by a conical jet model. Some combination of inhomogeneities and jet curvature or other direction change seem to better explain the observations. Of course the model that we have presented here is over simplified; in particular, even if the jet curves, it will presumably actually curve in 3-dimensions, making for a more complicated situation. However, more detailed models are beyond the scope of this paper.

\section{Conclusions}
\label{sec:11}
In the present work, we examined the long-term variability  of the blazar 3C 454.3 in optical, IR and radio bands for the extended period between February 2008 and April 2016.  This source showed significant variability on months to years timescales in all these bands. A long-term redder-when-brighter trend was observed in the (B--R) vs R and (V--R) vs R color indices. A bluer-when-brighter trend was observed in the optical band during the 2014 optical/IR flare similar to the trends seen during the 2007 flare \citep{Raiterietal2008}.  The radio spectrum remained fairly constant over a long period although we saw the spectral index increasing with flux during the 2008 radio flare.

There were tightly correlated variations in optical/IR bands with the radio bands lagging behind the optical bands by 15 to 100 days (depending on the segment). Strong correlations between the optical/IR bands with near zero lag suggest these emission regions are co-spatial. Optical and radio bands show correlations with time lags whose values are different in different years. This behavior can be incorporated in an inhomogeneous jet model where higher frequencies are emitted closer to the shock in the jet as compared to lower frequencies that are emitted further down the jet. As the lags are different during observing seasons or light curve segments it appears that the emitting regions change their orientation with respect to our line of sight.

\section{Acknowledgements}
\label{sec:12}
We thank the anonymous referee for extensive comments that substantially improved the manuscript. This research has made use of data from the OVRO 40-m monitoring program which is supported in part by NASA grants NNX08AW31G, NNX11A043G, and NNX14AQ89G and NSF grants AST-0808050 and AST-1109911. This publication makes use of data obtained at the Mets\"ahovi Radio Observatory, operated by the Aalto University. The program for calculating the DCF was developed by Edelson and Krolik, 1988, ApJ, 333, 646 for use on unevenly sampled and/or gapped data. An up-to-date SMARTS optical/near-infrared light curves are available at\\ \href{www.astro.yale.edu/smarts/glast/home.php}{www.astro.yale.edu/smarts/glast/home.php}. Data were also used from the updated archive of Steward Observatory available at\\ \href{http://james.as.arizona.edu/$\sim$psmith/Fermi/}{http://james.as.arizona.edu/$\sim$psmith/Fermi/}.

\bibliography{reference}
\bibliographystyle{aasjournal}
\end{document}